# Half-Duplex Active Eavesdropping in Fast Fading Channels:
# A Block-Markov Wyner Secrecy Encoding Scheme

George T. Amariucai, Member IEEE, and Shuangqing Wei, Member IEEE

*Abstract*—In this paper we study the problem of half-duplex active eavesdropping in fast fading channels. The active eavesdropper is a more powerful adversary than the classical eavesdropper. It can choose between two functional modes: eavesdropping the transmission between the legitimate parties (Ex mode), and jamming it (Jx mode) – the active eavesdropper cannot function in full duplex mode. We consider a conservative scenario, when the active eavesdropper can choose its strategy based on the legitimate transmitter-receiver pair's strategy – and thus the transmitter and legitimate receiver have to plan for the worst. We show that conventional physical-layer secrecy approaches perform poorly (if at all), and we introduce a novel encoding scheme, based on very limited and unsecured feedback – the *Block-Markov Wyner (BMW) encoding scheme* – which outperforms any schemes currently available.

## I. Introduction

A great number of recent works have been fueled by the still growing interest in physical layer secrecy. Most of them attempt to overcome the limitations of the classical wiretapper/eavesdropper scenarios of [1] or [2] (namely that no secret message can be successfully transmitted if the eavesdropper's channel is less noisy than the legitimate receiver's channel) by using some form of diversity.

The benefits of the ergodic-fading diversity upon the achievable secrecy rates have been exposed by works like [3], [4], [5] or [6]. A fast-fading eavesdropper channel is studied in [3] under the assumption that the main channel is a fixed-SNR additive white Gaussian noise (AWGN) channel. Although the secrecy capacity for fast-fading eavesdropper channels is still unknown, [3] provides achievable secrecy rates and shows that sometimes noise injection at the transmitter can improve these rates.

The different approach of [4] models both the main and the eavesdropper channels as ergodicly-fading AWGN channels. However, the fading is assumed to be slow enough to be considered constant for infinitely long blocks of transmitted symbols. The secrecy capacity is derived for this model, and the achievability part is proved by using separate channel encoding for each of the blocks. A similar approach is taken

G. Amariucai is with the Department of ECpE, Iowa State University. E-mail: gamari@iastate.edu.
S. Wei is with the Department of ECE, Louisiana State University. E-mail: swei@ece.lsu.edu.
This paper was supported in part by the Board of Regents of Louisiana under grants LEQSF(2004-08)-RD-A-17.

in [5] and [6], where the fading broadcast channel with confidential messages (BCC) is considered equivalent to a parallel AWGN BCC.

However, the slow fading ergodic channel model is quite restrictive. Although the model can be artificially created by a multiplexing/demultiplexing architecture as in [7], it still requires either coarse quantization or long delays (e.g. under fine quantization, for a channel state with low probability it may take a very large number of transmitted symbols to enable almost-error-free decoding).

With these considerations, we focus instead on a more practical scenario where both the main and the eavesdropper's channel are affected by *fast* stationary fading. However, unlike [3], we are concerned with a much stronger adversary: a half-duplex *active eavesdropper*.

In our channel model, depicted in Figure 1, the eavesdropper (Eve) has two options: either to jam the conversation between the legitimate transmitter (Alice) and the legitimate receiver (Bob) – Jx mode – or to eavesdrop – Ex mode – (our eavesdropper cannot function in full duplex mode, i.e. she cannot transmit and receive on the same frequency slot, at the same time). Both Alice and Eve (in Jx mode) are constrained by average (over each codeword) power budgets $\mathscr{P}$ and $\mathscr{J}$, respectively. Eve's purpose is to minimize the secrecy rate achievable by Alice, and to that extent she has to decide on the optimal alternation between the jamming mode and the eavesdropping mode. The state of each of the main and eavesdropper channels, i.e. the absolute squared channel coefficients (or simply "the channel coefficients" hence forth), which we denote by $h_M$ and $h_W$, respectively, are assumed to be available to the respective receivers. However, Bob does not know the exact state of Eve's channel, nor does Eve have any information about Bob's channel, except its statistics. In addition to fading, each channel is further distorted by an independent additive white complex Gaussian noise of variance $\sigma_N^2$. There exists a low-rate, unprotected (i.e. public) feedback channel between Bob and Alice.

The present paper is limited to the following simplifying (although not uncommon) assumptions.

i) Rayleigh fading: $h_M$ and $h_W$ are exponentially distributed, with parameters $\lambda_M$ and $\lambda_W$ respectively.

ii) The channel that links Eve (when in Jx mode) and Bob is error free and does not experience fading [8], [9].

iii) Eve only uses white Gaussian noise for jamming [10], [8], since this is the most harmful uncorrelated jamming strategy [11].

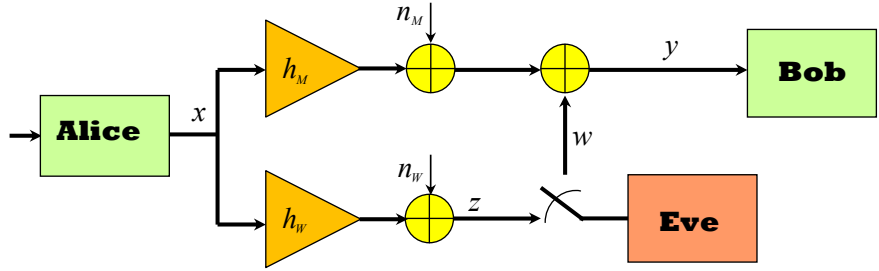

Fig. 1. Channel model

iv) Eve's exact jamming strategy (i.e. when and with what power she jams) is perfectly known to Bob (a posteriori) [1] so that Bob can employ coherent detection and communicate Eve's strategy to Alice, via the low-rate feedback link.

v) The instantaneous state of the main channel cannot be known to the transmitter Alice non-causally.

vi) The codewords are long enough such that not only the channel fading, but also the combination of channel fluctuation and Eve's alternation between jamming and eavesdropping display ergodic properties over the duration of a codeword.

vii) Eve employs an ergodic strategy, i.e. she uses the same statistics for alternating between Jx mode and Ex mode for every codeword.

viii) Eve has access to the exact value of $h_W$ only after she made her decision to eavesdrop (Ex mode), and has no information about the value(s) that $h_W$ might take while she is in Jx mode. This scenario models a situation where the training sequences, which are transmitted by Alice at a low rate, and are used by Bob to estimate the channel coefficient before the transmission of a block of symbols, are protected against eavesdropping (for instance, by using some form of secrecy encoding designed for non-coherent reception) or are simply unknown to Eve. Therefore, Eve cannot use the training sequences for estimating $h_W$. However, if Eve's channel is fading slowly enough, Eve may be able to perform some form of blind channel estimation. Nevertheless, this kind of procedure would require Eve to first listen to the incoming signal for a longer time interval, possibly as long as her channel coherence time. Under these circumstances, Eve has to take the decision on whether to jam or eavesdrop in the absence of any non-causal channel state information (i.e. randomly).

Our contributions can be stated as follows:
- We introduce the concept of (half-duplex) active eavesdropper;
- We show that, under our conservative scenario, Wyner's scheme [1] performs poorly (if at all);

[1]To estimate exactly where Eve jams, one may argue that Bob needs the coherence time of the channel (which includes Eve's alternation between jamming and eavesdropping) to span several channel uses. But since the jammer's hardware construction will most probably prevent Eve from switching between Ex and Jx modes instantaneously, the coherence time of the jamming is not likely to pose any problems. On the other hand, the coherence time of the channel coefficient $h_W$ can be assumed large enough to allow for Bob's binary hypothesis testing (Ex or Jx) without bringing up any of the problems of [4], [5], [6], where the coherence time needs to be large enough to allow hypothesis testing (decoding) between a number of hypotheses (codewords) which increases exponentially with the codeword length.

- We provide a novel block-Markov Wyner (BMW) secrecy encoding scheme, which requires a low-rate, unsecured feedback link from Bob to Alice, and can improve the secrecy rate significantly;
- Our BMW scheme employs a diversity of concepts, such as a-posteriori Wyner-type binning, block-Markov secrecy encoding and encoding for a compound channel;
- We provide a secrecy-encoding method for multiple-access channels (MACs), where even if some of the users are not decodable by the receiver, they can still help with the transmission of secrecy.

We should note that our BMW scheme displays greater generality, and is not limited to the present scenario. In fact, we are currently investigating its use in extending the concepts of physical-layer secrecy to more realistic slowly-fading-channel models.

## II. THE CONSERVATIVE SCENARIO AND THE ACHIEVABLE SECRECY RATES

Physical-layer secrecy is synonymous to Wyner-type secrecy-encoding schemes [1]. The main idea behind these schemes is to create a special channel code, taylored to exploit the physical disadvantages of Eve's channel. Alice and Bob agree on a certain binning strategy, taylored to a certain pair of one main and one eavesdropper channels. Nevertheless, Eve's actual channel quality remains unknown. In fact, in most passive-eavesdropper scenarios, Eve herself remains undetected. Therefore, the Wyner-type schemes are not influenced by Eve's position, but by the legitimate parties' perception about her position.

In our active-eavesdropper scenario, Eve can alternate between jamming and eavesdropping. However, it is not this feature that turns out to be devastating for the transmission of secret messages, as much as the legitimate parties' uncertainty about Eve's strategy. To provide a stable framework for our investigation, we shall agree on the following notation and concepts. Throughout this paper, we shall denote $q = Pr\{\text{Ex mode}\}$ the probability that Eve is in Ex mode over a given frame. Note that under our assumptions, $q$ uniquely determines Eve's strategy over a frame, and remains unknown to Alice until the end of the frame.

Obviously, Eve's presence causes a modification of the channel statistics (as Alice and Bob see them). For example, whenever Eve is eavesdropping (in Ex mode), the main channel instantaneous SNR is $\frac{h_M P}{\sigma_N^2}$, while the SNR of Eve's channel is $\frac{h_W P}{\sigma_N^2}$ – no modification here. However, when Eve is jamming (in Jx mode), the main channel instantaneous SNR is $\frac{h_M P}{\sigma_N^2 + J}$, where $J$ is the instantaneous jamming power subject to the constraint $\mathbf{E} J \leq \mathscr{J}$, while the SNR of Eve's channel is zero (recall that whenever Eve jams, she cannot simultaneously listen on the same frequency slot).

From Alice's and Bob's perspective, the new *equivalent* channel coefficients can be written as

$$\widetilde{h_M} = \begin{cases} h_M & \text{if Ex mode} \\ \frac{h_M \sigma_N^2}{\sigma_N^2 + J} & \text{if Jx mode} \end{cases} \quad (1)$$

and

$$\widetilde{h_W} = \begin{cases} h_W & \text{if Ex mode} \\ 0 & \text{if Jx mode,} \end{cases} \quad (2)$$

with the observation that Alice and Bob still have to agree in advance on how long Eve should be considered in Ex mode, and how long she should be considered in Jx mode (i.e. on the value of $q$). This kind of information will determine the encoding strategy and the achievable secrecy rate.

Denote by $X$ the random variable at the input of the two channels, and by $Y$ and $Z$ the corresponding random variables received by Bob and Eve, respectively. According to [2], the secrecy capacity of our model (under the assumption that the equivalent channel coefficients $\widetilde{H_M}$ and $\widetilde{H_W}$ become perfectly known to Bob and Eve during transmission, and hence can be considered as outputs of the channel) is given by

$$C_s = \max_{V \to X \to YZ} \left[ I(V; Y, \widetilde{H_M}) - I(V; Z, \widetilde{H_W}) \right] \geq$$
$$\geq \max_{V \to X \to YZ} \left[ I(V; Y | \widetilde{H_M}) - I(V; Z | \widetilde{H_W}) \right], \quad (3)$$

where the maximization is over all joint probability distributions of $V$ and $X$ such that $V \to X \to YZ$ form a Markov chain. The inequality in (3) follows from the independence between $V$ and $H_W$, and holds with equality if $V$ is also independent of $H_M$ (i.e. Alice has no a-priori channel state information – CSI). Since the optimal choice of $V$ and $X$ is presently unknown, we shall henceforth concentrate on the *achievable secrecy rate* (instead of secrecy capacity) obtained by setting $V = X$ and picking a complex Gaussian distribution for $X$, with zero mean and variance $P$. Under these constraints, the achievable secrecy rate (over a frame) becomes:

$$R_s = \mathbf{E}_{\widetilde{h_M}, P} \left[ \log(1 + \frac{\widetilde{h_M} P}{\sigma_N^2}) \right] -$$
$$- \mathbf{E}_{\widetilde{h_W}, P} \left[ \log(1 + \frac{\widetilde{h_W} P}{\sigma_N^2}) \right], \quad (4)$$

where $P$ is the instantaneous transmitter power and is subject to the constraint $\mathbf{E} P \leq \mathscr{P}$.

As we have mentioned earlier, a classical Wyner-type secrecy-encoding scheme uses a codebook designed beforehand, and taylored to the specific channel conditions (assumed known in advance). If a codebook were designed for a specific parameter $q = q_0$, it would fail if Eve decided to use any different strategy. More precisely, if Eve used $q_1 > q_0$, the perfect secrecy of the message would be compromised (we call this *secrecy outage*), while if Eve used $q_2 < q_0$, the secret message would become unintelligible to Bob (we call this *intelligibility outage*).

As a result, the legitimate parties have to use a transmission strategy that can protect both the secrecy and the intelligibility of the secret message, under any strategy that Eve might use. This problem is best modeled by the conservative scenario that makes the assumption that Eve knows Alice's strategy in advance. Results for the best-case scenario (or the *minimax scenario*), where Alice and Bob know Eve's strategy in advance, are given in [12]. Although those results have less practical value, they can function as an upper-bound for the achievable secrecy rate, and will be used for comparison in the numerical results section.

The simplest encoding scheme that offers secrecy protection in our conservative scenario is one of Wyner type, with a forwarding rate low enough to protect the message against the most powerful attempt to induce intelligibility outage (i.e. when Eve is in Jx mode all the time), and with a secrecy rate low enough to offer protection against the most powerful attempt to induce secrecy outage (i.e. when Eve is in Ex mode all the time). The achievable secrecy rate for this kind of scheme is

$$R_{s,wcs} = \left[ \mathbf{E}_{h_M, P, J} \left[ \log(1 + \frac{h_M P}{\sigma_N^2 + J}) \right] - \right.$$
$$\left. - \mathbf{E}_{h_W, P} \left[ \log(1 + \frac{h_W P}{\sigma_N^2}) \right] \right]^+ \quad (5)$$

(the subscript "wcs" stands for "worst-case scenario"), and is achieved under the *equivalent* channel coefficients $\widetilde{h_M} = h_M \frac{\sigma_N^2}{\sigma_N^2 + J}$ and $\widetilde{h_W} = h_W$. The following two propositions show that randomizing the instantaneous power is not a geed idea for either Alice ($P$) or Eve ($J$).

*Proposition 1:* When no channel state information is available to the transmitter, the optimal transmitter strategy is to allocate constant power $P = \mathscr{P}$ to each symbol.

*Proof:* Recall our assumption that both $h_M$ and $h_W$ are exponentially distributed, with parameters $\lambda_M$ and $\lambda_W$, respectively. This means that $\widetilde{h_M}$ is also exponentially distributed, with parameter $\lambda_M (1 + \frac{\mathscr{J}}{\sigma_N^2})$. Denote the probability distribution of $\widetilde{h_M}$ by $f_M(x) = \lambda_M e^{-\lambda_M x}$, and of $\widetilde{h_W} = h_W$ by $f_W(x) = \lambda_W e^{-\lambda_W x}$.

If $\lambda_M (1 + \frac{\mathscr{J}}{\sigma_N^2}) \geq \lambda_W$ (Eve's equivalent channel is statistically "better"), then the achievable secrecy rate is zero. In this case the way Alice distributes her power (without knowledge of the exact channel coefficients) is irrelevant, and a constant power allocation is as good as any. Hence we shall concentrate on the case when $\lambda_M (1 + \frac{\mathscr{J}}{\sigma_N^2}) < \lambda_W$.

We need to prove that for this case, the function

$$R_{s,wcs}(P) = \mathbf{E}_{h_M, J} \log(1 + \frac{\widetilde{h_M} P}{\sigma_N^2}) -$$
$$- \mathbf{E}_{h_W} \log(1 + \frac{h_W P}{\sigma_N^2}) \quad (6)$$

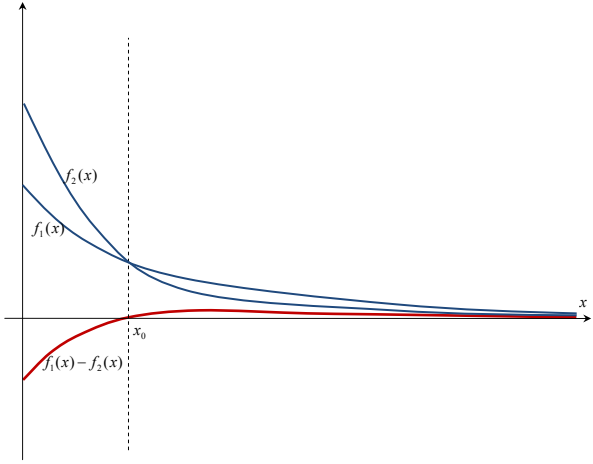

Fig. 2. Exponential distributions and their difference.

is a concave $\cap$ function of $P$. We can write

$$R_{s,wcs}(P) = \int_0^\infty \log(1 + \frac{xP}{\sigma_N^2})(f_M(x) - f_W(x))dx. \quad (7)$$

Note that $f_M(x) - f_W(x)$ is negative for $x \in [0, x_0)$ and positive for $x \in [x_0, \infty)$, where $x_0$ is the (unique) solution of $f_M(x) = f_W(x)$. Moreover, $\int_0^\infty f_M(x)dx = \int_0^\infty f_W(x)dx = 1$, which results in

$$\int_0^{x_0} [f_W(x) - f_M(x)]dx = \int_{x_0}^\infty [f_M(x) - f_W(x)]dx. \quad (8)$$

A graphical representation of these functions is given in Figure 2, where we used the notation $f_1 = f_M$ and $f_2 = f_W$. Take an *increasing function* $G(x)$. We can write

$$\int_0^{x_0} G(x)[f_W(x) - f_M(x)]dx \leq$$
$$\leq \int_0^{x_0} G(x_0)[f_W(x) - f_M(x)]dx =$$
$$= \int_{x_0}^\infty G(x_0)[f_M(x) - f_W(x)]dx \leq$$
$$\leq \int_{x_0}^\infty G(x)[f_M(x) - f_W(x)]dx. \quad (9)$$

Now, taking $G(x) = \log(1 + \frac{xP}{\sigma_N^2})$ we see that $R_{s,wcs}(P)$ is a positive function of $P$; taking $G(x) = \frac{dF(P)}{dP} = \frac{x}{\sigma_N^2 + xP}$ we see $R_{s,wcs}(P)$ is increasing; and taking $G(x) = \frac{d^2F(P)}{dP^2} = -\left(\frac{x}{\sigma_N^2 + xP}\right)^2$ we see that $R_{s,wcs}(P)$ is concave. ∎

*Proposition 2:* When in jamming (Jx) mode, Eve's optimal strategy is to use the same jamming power $J = \frac{\mathscr{J}}{1-q}$ across all channel realizations involved.

*Proof:* The result follows directly from (5), where only the first term depends on $J$, and that term is a convex function of $J$. ∎

As a result, the achievable worst-case-scenario secrecy rate is now simply

$$R_{s,wcs} = \left[\mathbf{E}_{h_M}\left[\log(1 + \frac{h_M \mathscr{P}}{\sigma_N^2 + \frac{\mathscr{J}}{1-q}})\right] - \mathbf{E}_{h_W}\left[\log(1 + \frac{h_W \mathscr{P}}{\sigma_N^2})\right]\right]^+ \quad (10)$$

and is rarely strictly positive (if and only if $\lambda_W > \lambda_M(1 + \frac{\mathscr{J}}{\sigma_N^2})$). For a large jamming-power-to-noise ratio $\mathscr{J}/\sigma_N^2$, this implies that Eve's physical channel needs to be impractically worse than Bob's.

However, the above scheme does not take full advantage of the model characteristics. Recall the original assumption that Eve can function only as a half-duplex terminal. Therefore, whenever Eve is in Jx mode, she cannot eavesdrop – so the whole transmission remains perfectly secret to Eve – and conversely, if she is in Ex mode, Eve cannot simultaneously jam the transmission.

In the next section we develop an alternative transmission scheme, which greatly improves the achievable secrecy rate, and is tuned to specifically exploit the active eavesdropper's limitations. More specifically, we "quantize" the interval $[0, 1]$ to which Eve's strategy $q$ belongs into several smaller intervals, and we design an encoding/decoding strategy such that the worst-case scenario outlined above is only applied on one of these sub-intervals. The finer the resolution, the smaller the loss of secrecy rate due to the uncertainty about Eve's strategy. To make this encoding strategy work, we use the following techniques: (i) a posteriori Wyner-type binning, (ii) secret key generation, (iii) block-Markov secrecy encoding and (iv) encoding for a compound channel.

## III. THE BLOCK-MARKOV WYNER (BMW) ENCODING SCHEME

There are two main reasons why Wyner's scheme [1] does not work in our model. First, Alice does not know the statistics of Bob's channel in advance – Eve has control over the signal-to-noise ratio of this channel. Therefore, the main channel can be modeled as a compound channel. In order to reliably transmit a message to Bob, Alice should use a special kind of encoding. It was shown in [13] that the same layered encoding technique that achieves the points on the boundary of the capacity region for broadcast channels can also be used for transmission over compound channels. Our scheme uses the broadcast layered encoding of [13] to ensure that reliable transmission is possible between Alice and Bob even in the most unfavorable conditions. However, even if such a scheme is used, Alice cannot know in advance which messages will actually be decodable by Bob.

The second reason is that Alice does not know the statistics of Eve's channel in advance – due to the alternation between jamming and eavesdropping, Eve's equivalent channel is actually weaker than her physical channel. Therefore, Alice cannot directly transmit a secret message at a rate larger than $R_{s,wcs}$ in (10), because she is not sure whether the secrecy would be compromised or not.

We solve these two problems by introducing a posterior-binning Wyner-type encoding scheme. Instead of transmitting a secret message by Wyner's scheme, we choose to transmit white noise, and agree on a secret key at the end of transmission, once the channel quality becomes available (a posteriori). The secret key is then used over the next transmission interval, to encrypt a secret message, which is then transmitted at the same time with another sequence of white noise from which a new secret key is distilled, and so on. Our approach is a sequential one, and requires that Bob should be actively involved in the secrecy encoding process. Bob's involvement consists of estimating and feeding back to Alice the exact value of Eve's strategy $q$. The detailed description is given below. However, before we get to that, we need to present some considerations on Wyner's original encoding scheme [1], which will help build some intuition regarding the principle of our own scheme.

### A. An alternative to Wyner's secrecy encoding scheme for regular passive-eavesdropper channels: a posteriori binning

We begin this discussion by studying a scenario where, before the transmission takes place, Alice and Bob already share a secret key (perhaps one that was agreed upon after the previous transmission). Then in addition to the secret message that can be encoded by Wyner's scheme, another secret message can be transmitted over the channel. This latter message is encrypted using the secret key. We provide two encoding schemes that can both achieve the simultaneous transmission of the two secret messages.

Denote the capacities of the channels from Alice to Bob and from Alice to Eve by $C_M$ and $C_E$, respectively, the achievable secrecy rate (under Wyner's original scheme) by $R_k$, the rate of the encrypted message by $R_s$ and the codeword length by $N$.

*Scheme 1: Wyner's scheme with an encrypted message.* Alice bins the codebook (containing $2^{NC_M}$ codewords) into $2^{NR_k}$ "super-bins", such that $R_k \leq C_M - C_E$. The first secret message picks the index of a super-bin. The super-bin is then binned again into $2^{N(C_M-R_s-R_k)}$ bins (each containing $2^{NR_s}$ bin-words). One of the bins is picked randomly, while a specific codeword in that bin is picked according to the encrypted message.

*Scheme 2: The alternative encoding scheme.* The codebook is randomly binned into $2^{N(C_M-R_s)}$ bins – let us denote these as "pre-bins". Each pre-bin consists of $2^{NR_s}$ bin-words. The bins are then randomly grouped into $2^{NR_k}$ "super-bins", such that each super-bin consists of $2^{N(C_M-R_s-R_k)}$ bins, and where $R_k$ is picked to satisfy $R_k \leq C_M - C_E$. The first secret message picks the index of a super-bin. A bin inside that super-bin is randomly picked, and the transmitted codeword is then picked by the encrypted message inside this bin.

The two schemes are equivalent, and they are described in Figure 3. However, as we shall see shortly, the applicability of *Scheme 2* is more relevant to our compound channel scenario. We should recall here that Wyner's original encoding scheme [1] involves a random binning of the codebook into bins which are, each of them, good codes for Eve's channel.

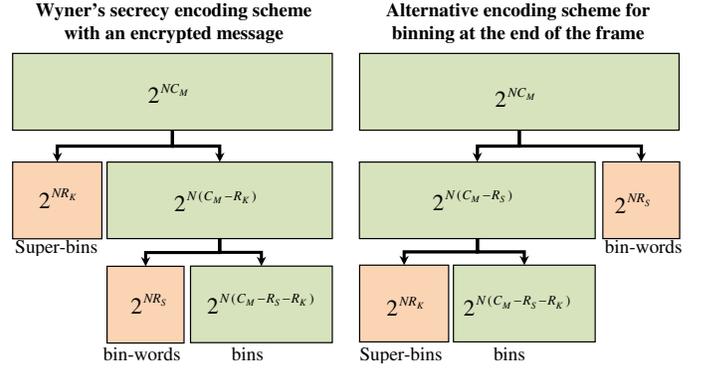

Fig. 3. Alternative binning: Wyner's secrecy encoding scheme with an additional encrypted message, and the basis of our block-Markov Wyner encoding scheme.

The actual transmission does not contain any information about the binning itself. Hence, the same "random" binning needs to be done separately at Alice (before the transmission takes place) and at Bob (before he can begin decoding). The reason why Alice performs the binning of the codebook before transmitting is because she needs to send a *meaningful* secret message over the coming frame. Therefore, the transmitted codeword needs to belong to the particular bin indexed by the secret message.

This suggests that if the "secret message" transmitted by Alice had no meaning (i.e. if Alice picked this message in a random fashion), both Alice and Bob could perform the binning of the codebook after the transmission ends. The "secret message" generated this way could be thought of as a *secret key* for encrypting a meaningful message over the next shared frame.

Suppose that Eve's channel is unknown to Alice and Bob until the transmission of the current codeword ends. The first transmitted codeword is randomly selected from the whole unbinned codebook. After the transmission ends, Alice and Bob realize that the secrecy capacity was $R_s$. Both Bob and Alice can now proceed to the (same) binning of the codebook. As a result, the same single bin will be identified by both legitimate parties as containing the transmitted message, and its index will be secret to Eve. Clearly, the secret message conveyed by the index of this bin has no meaning. Nevertheless, it can be used over the next frame, as a secret key. Over the second frame, Alice and Bob use *Scheme 2* above. The codebook is randomly binned before transmission, into $2^{N(C_M-R_s)}$ bins that could each be regarded as a code for carrying the encrypted message. One of the bins will be selected randomly, and the encrypted message will select the exact codeword to be transmitted. This method of transmission ensures that the encrypted message does not overlap with the secret key that needs to be generated at the end of the frame – the encrypted message has nothing to do with how the bins are ultimately chosen, as seen in Figure 3. The encrypted message may be *decodable, but not decryptable* by Eve. After the transmission of the second frame takes place, Alice and Bob realize that the secrecy capacity was $R_k$. The indices of the bins are "randomly" grouped by both Alice and Bob into $2^{NR_k}$ super-bins, and a new secret key is agreed upon by the legitimate parties. The protocol continues in the same manner.

Three observations are in order. First, the secret key (decided upon at the end of the frame) and the encrypted message (carried by the frame) cannot overlap and maintain the same equivocation at Eve – see the *one-time pad* [14]. Hence, in the above description of the protocol, it is required that $R_s + R_k \leq C_M$. Second, note that $R_s = R_k$ (since the key is used as a one-time pad to encrypt the secret message of the next frame), therefore, if our previous condition holds in the form $R_s < C_M/2$, the transmission of the meaningful secret message can be done at almost the secrecy capacity, with a small initial penalty (due to the fact that the first frame does not carry an encrypted message) which becomes negligible as the number of transmitted frames increases.[2] Third, our new protocol can be used whenever Alice does not have a good description of Eve's channel over a frame until the transmission of the corresponding codeword ends, which is precisely the case with our current model.

### B. Detailed description of the BMW encoding scheme

At this point, we restrict our analysis to particular frame (we shall denote the span of a codeword by "frame"). How Eve should deal with different frames will be discussed in Theorem 5. Over this frame, we assume that Eve chooses an arbitrary strategy $q = Pr\{\text{Ex mode}\}$. Once the transmission of the codeword is finished, Bob can accurately evaluate the parameter $q$. Bob can then feed this value back to Alice. Note that the knowledge of $q$ provides Alice with the statistical description of both the main channel – determined by the jamming probability $(1-q)$ – and the eavesdropper's channel – determined by the eavesdropping probability $q$. Before learning Eve's strategy, the channel between Alice and Bob appears like a compound channel to the legitimate parties. The possible states of this channel are given by the possible values of Eve's strategy $q$, which belongs to the interval $[0, 1]$. To transform this uncountable set of possible channel states into a finite set, we divide the interval $[0, 1]$ to which $q$ belongs into $n$ subintervals such that

$$[0,1] = [q_0, q_1) \cup [q_1, q_2) \ldots \cup [q_{n-1}, q_n] \quad (11)$$

where $q_0 = 0$ and $q_n = 1$.

For conveying a message to Bob, Alice uses an $n$-level broadcast-channel-type codebook, as in [13]. Level $i$ is allocated power $(1 - \alpha_i)\alpha_{i-1} \ldots \alpha_1 \mathscr{P}$ (with $\alpha_j \in [0, 1]$ $\forall j = 1, \ldots, n-1$ and $\alpha_n = 0$) and is designed to deal with a jammer which is on with probability $1 - q_{i-1}$ over each channel use. Also note that $q_0 < q_1 < \ldots < q_n$. In the remainder of this paper, we shall say that level $i$ is "stronger" than level $j$ if $i < j$, i.e. if level $i$ can deal with a jammer which is on more often. The notation is fully justified by Lemma 3 below.

Denote the rates of the different encoding levels as:

$$R_1 = \mathbf{E}_{h_M}\left[\log\left(1 + \frac{(1-\alpha_1)\mathscr{P}h_M}{\sigma_N^2 + \alpha_1 \mathscr{P} h_M + \mathscr{J}}\right)\right] \quad (12)$$

for the strongest level, which can deal with the case when Eve is always in Jx mode, i.e. $q = q_0 = 0$,

$$R_i = \mathbf{E}_{h_M}\bigg[q_{i-1}\log\bigg(1+$$
$$+\frac{(1-\alpha_i)\alpha_{i-1}\ldots\alpha_1\mathscr{P}h_M}{\sigma_N^2 + \alpha_i\ldots\alpha_1 \mathscr{P} h_M}\bigg) + (1-q_{i-1})\log\bigg(1+$$
$$+\frac{(1-\alpha_i)\alpha_{i-1}\ldots\alpha_1\mathscr{P}h_M}{\sigma_N^2 + \alpha_i\ldots\alpha_1 \mathscr{P} h_M + \frac{\mathscr{J}}{1-q_{i-1}}}\bigg)\bigg], \quad (13)$$

for $i = 2, 3 \ldots n-1$, and finally

$$R_n = \mathbf{E}_{h_M}\bigg[q_{n-1}\log\bigg(1 + \frac{\alpha_{n-1}\ldots\alpha_1\mathscr{P}h_M}{\sigma_N^2}\bigg) +$$
$$+(1-q_{n-1})\log\bigg(1 + \frac{\alpha_{n-1}\ldots\alpha_1\mathscr{P}h_M}{\sigma_N^2 + \frac{\mathscr{J}}{1-q_{n-1}}}\bigg)\bigg], \quad (14)$$

for the weakest level, corresponding to the case when Eve is in Jx mode with probability $1 - q_{n-1}$. Note that the encoding levels are designed such that Bob decodes the stronger levels first, and treats the remaining un-decoded messages as white noise. The codebook for level $i$ contains $2^{NR_i}$ codewords of length $N$, generated such that each component of each codeword represents an independent realization of a Gaussian random variable of mean 0 and variance $(1-\alpha_i)\alpha_{i-1}\ldots\alpha_1\mathscr{P}$, where $\alpha_n = 0$ for compatibility.

Also note that our scheme uses a constant transmit power $P = \mathscr{P}$ over the whole frame. Although randomizing the transmit power may sometimes improve the achievable secrecy rate, this study is beyond the scope of our paper. However, we already know that Eve's optimal strategy is to use a constant jamming power $\mathscr{J}$ on all jammed channel uses. This is because all the rates $R_1, \ldots, R_n$ defined above are convex functions of $\mathscr{J}$ (see Proposition 2).

The relative strength of the encoding levels is established by the following lemma.

*Lemma 3:* If Eve uses a parameter $q \in [q_{i-1}, q_i)$ over a frame, then the messages encoded in levels $1, 2, \ldots, i$ are intelligible by Bob at the end of the frame. Thus the forwarding rate from Alice to Bob is $R_{M,i} = R_1 + R_2 + \ldots + R_i$.

*Proof:* In order to prove that the encoding levels with lower indices are stronger in the sense that they can deal with a worse jamming situation, it is enough to show that $R_i$ as defined in (13) is an increasing function of $q$. In other words, encoding level $i$, transmitting at a rate $R_i$, is intelligible by Bob whenever Eve is in jamming mode with a probability less than $(1 - q_{i-1})$. But this is a direct consequence of Lemma 7 in Appendix A, if we simply replace $x$ by $\frac{(1-\alpha_i)\alpha_{i-1}\ldots\alpha_1\mathscr{P}h_M}{\sigma_N^2 + \alpha_i\ldots\alpha_1 \mathscr{P} h_M}$ and $y$ by $\frac{\mathscr{J}}{\sigma_N^2 + \alpha_i\ldots\alpha_1 \mathscr{P} h_M}$ (see Appendix A). ∎

---

[2]Assume that Eve's channel conditions are always the same. As an example, consider a codebook with 10000 codewords, which is used for transmitting a secret message of length $\log(50)$ bits, according to our protocol. Take any random frame. For transmitting the encrypted message, the codebook is binned into 200 bins, each containing 50 codewords. One of the bins is picked randomly, and the encrypted message picks one of the 50 codewords in the bin. After the transmission takes place, Alice and Bob both group the original 200 bins into 50 "super-bins" (each containing 4 original bins), using the same "recipe". The secret key is the index of the super-bin to which the transmitted codeword belongs. Note that the actual codeword that was transmitted inside this super-bin is picked independently of the choice of the super-bin.

Consider the first frame, for which the transmitted message carries no useful information, but rather its symbols are selected in a random, i.i.d. fashion. Once Alice receives the feedback sequence from Bob at the end of the frame, describing Eve's strategy (i.e. the value of $q$ – actually, as we shall see shortly, only the interval $[q_{i-1}, q_i)$ that contains $q$ is enough information for Alice, thus the length of the feedback sequence need not be larger than $\log(n)$), Alice and Bob can separately agree on the same secret message, as described in the protocol above. This message will function as a secret key for encrypting a meaningful secret message over the next frame. In turn, the secret message agreed upon at the end of the second frame can function as a secret key for the third frame, and so on.

To formalize the intuitive description above, we begin by stating several definitions:

- The "encrypted message" is a meaningful secret message, encrypted with the help of a secret key that was generated in the previous frame.
- The "secret key" is a meaningless random message, which is perfectly secret to Eve, is agreed upon by both Alice and Bob at the end of the frame, and can be used for the encryption of a secret message (of at most the same length) over the next frame.
- The term "secret key rate" refers to the rate at which a secret key is generated at the end of a frame – the correspondent of Wyner's "secrecy capacity".
- The term "achievable secrecy rate" refers to the rate of transmission of the encrypted message.

Our encoding scheme works as follows. First, the $n$ codebooks, indexed by $i$, with $i \in \{1, 2, \ldots, n\}$ are generated as described above, and are made available to all parties. On a given frame, Alice transmits an encrypted message, at a rate

$$R_s \leq 0.5 R_1 \tag{15}$$

(this constraint is a result of planning ahead for Eve's most destructive behavior, and we show in Theorem 5 below that it does not incur any loss of performance under Eve's optimal strategy) – note that the *encrypted message* is encrypted with the help of a secret key generated over a previous frame. To transmit the encrypted message, Alice randomly bins codebook 1 into $2^{N(R_1 - R_s)}$ bins. One of the bins (each containing $2^{NR_s}$ codewords) will be picked randomly (uniformly), and the encrypted message will pick a codeword from this bin for transmission. Recall that the reason why Alice cannot directly bin the codebook for generating the secret key is because Eve's strategy (hence her equivalent channel) is unknown until the end of the frame. An additional $n-1$ codewords are also chosen randomly, one from each of the remaining $n-1$ codebooks of rates $R_2, R_3, \ldots, R_n$. Alice's transmitted sequence is the sum of the $n$ codewords.

At the end of the frame, Bob feeds back to Alice the exact value of Eve's strategy $q$ over that frame. In order to agree on a secret key, Alice and Bob first need to know which encoding levels are decodable by Bob, and which are decodable by Eve. Only the information encoded in those levels that are decodable by Bob, but are not perfectly decodable by Eve, can contribute to the generation of the secret key.

Due to the construction of the code (see Lemma 3), it is clear that under any jamming/ eavesdropping strategy, Bob will be able to decode the strongest level first, treating the other levels as white noise, and then perform successive interference cancellation to decode increasingly weaker levels. However, the same statement cannot be made for Eve. Eve's channel is quite different from Bob's. While the code is designed to handle Bob's unknown-length interference channel, Eve sees an interference-free channel that is totally interrupted $(1-q)$ of the time. In the general case, it is thus possible that the order of strength of the encoding levels, from Eve's perspective, is not the same as that from Bob's perspective. For example, for a code with 7 levels Bob might be able to decode only levels $1, 2, 3, 4$, while Eve-A may be able to perfectly decode only levels $1, 4, 6, 7$. In this case, we can re-order the levels from Eve-A's perspective, as $1, 4, 6, 7, 2, 3, 5$. The first four levels are decodable by Eve-A perfectly, the next two are decodable by Bob, but not by Eve-A, and the last level is decodable by neither. Only levels 2 and 3 can be used for generating the secret key.

For the general case, we shall denote the ordered set of indices corresponding to the encoding levels specified by their rates in (12)-(14) by $\mathscr{I}$, and the set of indices corresponding to the order of strength of the encoding levels from Eve's perspective by $\widehat{\mathscr{I}}$. There exists a bijection (i.e. a re-ordering) $\mathbb{B} : \mathscr{I} \to \widehat{\mathscr{I}}$, defined as follows: (1) the set of indices (in arbitrary order) corresponding to levels that are perfectly decodable by Eve is denoted $\mathscr{I}_e$; (2) the set of indices (in arbitrary order) corresponding to levels that are not perfectly decodable by Eve, but perfectly decodable by Bob is denoted $\mathscr{I}_k$; (3) the set of indices (in arbitrary order) corresponding to levels that are not perfectly decodable by either Eve or Bob is denoted $\mathscr{I}_n$; (4) the ordered set $\widehat{\mathscr{I}}$ is defined as

$$\widehat{\mathscr{I}} = \{\mathscr{I}_e, \mathscr{I}_k, \mathscr{I}_n\}. \tag{16}$$

Furthermore, we define $\mathscr{I}_{ne} = \{\mathscr{I}_k, \mathscr{I}_n\}$ as the set of indices corresponding to encoding levels which are not perfectly decodable by Eve. The method of encoding is described in Figure 4. Theorem 4 below provides the achievable secret key rate for the general case.

*Theorem 4:* Consider a given quantization $\{q_i | i \in \widehat{\mathscr{I}}\}$ of the interval $[0, 1]$, and a given power allocation between the corresponding encoding levels $\{\alpha_i | i \in \widehat{\mathscr{I}}\}$. Suppose that Eve picks a strategy $q \in [q_{i-1}, q_i)$ over a frame. Then the following secret key rate is achievable over that frame (where the key is generated at the end of the frame):

$$R_{k,i} = \sum_{j \in \mathscr{I}_k} [R_j - R_{E,j}], \tag{17}$$

where:

- $R_j$ are defined as in (12)-(14) for $j = 1, 2, \ldots, n$,
- $R_{E,j}$, $j \in \mathscr{I}_{ne}$ are selected such that they satisfy the following set of conditions:

$$R_{E,1} \geq 0.5 R_1 \text{ if } 1 \in \mathscr{I}_{ne}, \tag{18}$$

(this condition states that the secret key rate assigned to the first encoding level should not exceed $0.5 R_1$, because otherwise Eve might choose $q = q_1$, and the achievable secrecy rate could not exceed $R_1 - R_k$ – see Theorem 5)

$$R_{E,l} \leq R_l \; \forall l \in \mathscr{I}_{ne} \quad (19)$$

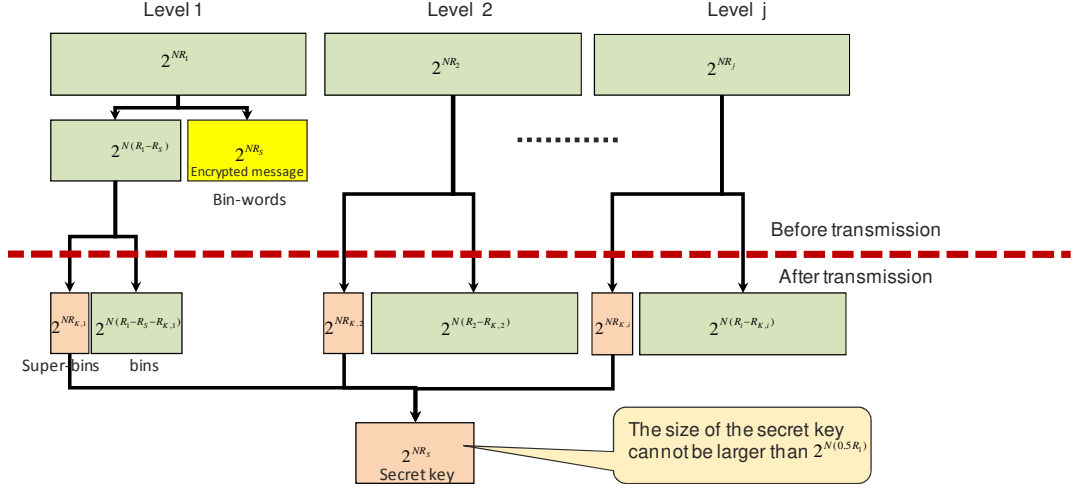

Fig. 4. BMW encoding method – most general case, when $1 \in \mathscr{I}_{ne}$.

(this condition ensures the feasibility of encoding),

$$\sum_{l \in \mathscr{S}} R_{E,l} \leq q\mathbf{E}_{h_W}\left[\log\left(1 + \frac{\sum_{l \in \mathscr{S}}(1-\alpha_l)\alpha_{l-1}\ldots\alpha_1 \mathscr{P} h_W}{\sigma_N^2}\right)\right] - \epsilon, \quad (20)$$

for any subset $\mathscr{S}$ of $\mathscr{I}_{ne}$, and

$$\sum_{l \in \mathscr{I}_{ne}} R_{E,l} = q\mathbf{E}_{h_W}\left[\log\left(1 + \frac{\sum_{l \in \mathscr{I}_{ne}}(1-\alpha_l)\alpha_{l-1}\ldots\alpha_1 \mathscr{P} h_W}{\sigma_N^2}\right)\right] - \epsilon, \quad (21)$$

with $\epsilon$ positive and arbitrarily close to zero (the latter two conditions ensure the secrecy of the key).

The expressions in (20) and (21) use the convention $\alpha_n = 0$. Note that the bijection $\mathbb{B}$ defined above also depends on Eve's strategy $q$, and hence on the interval $i$ to which $q$ belongs. Therefore, the set of indices $\mathscr{I}_k$ depends on $i$.

The following observations are in order.

(1) Sometimes (when Eve's channel is much worse than Bob's) the system of inequalities in (18) - (21) may have no feasible solution. Under these circumstances, one possible approach is to find a solution of the inequalities in (19) - (21) (i.e. to ignore the condition in (18)). This would distill a secret key at a rate larger than $0.5R_1$. Nevertheless, the entire secret key may be used by time sharing. For example, for $\mathscr{I}_{ne} = \{1\}$, we can take $R_{E,1} = 0.25R_1$ for two consecutive frames (the first level would generate $2 \times 0.75NR_1 = 1.5NR_1$ secret key bits, and transmit only $2 \times 0.25NR_1 = 0.5NR_1$ bits of the secret message), for the third frame we can use the entire level 1 (at rate $R_1$) to transmit an encrypted message (of length $1.5NR_1 - 0.5NR_1 = NR_1$ bits). With this observation, in the remainder of this work we shall only focus on the cases in which (18) - (21) admit feasible solutions.

(2) Since the fundamentals of our approach to the proof of this theorem reside in Wyner's original results [1], we are currently constrained to involving all users in $\mathscr{I}_{ne}$ in the generation of the secret key. That is, we are subject to the constraint in (21), and its essentiality will be reflected in the proof. Although a larger secrecy rate might be achieved by involving only a proper subset of $\mathscr{I}_{ne}$ in the generation of the secret key, this kind of improvement is beyond the purpose of this paper, and will be considered for further research.

(3) Although similar-flavor results have been obtained in [15], [16], our results are quite different because they involve "users" which are not decodable by Bob. A discussion of the issues addressed in [15], [16] is provided in Appendix B.

*Proof:* The proof is based on two observations. First, we have already shown that if the secret message is not a meaningful one, the binning of Wyner's scheme can be done at the end of the transmission, when the statistical properties of Eve's channel are known to both Bob and (through feedback) to Alice. To accomplish this, both Alice and Bob will have to memorize a set of "binning recipes", one for each possible value of Eve's strategy (actually only the interval $[q_{i-1}, q_i)$ to which $q$ belongs, and not the exact value of $q$, matters in our case). This is a bit different from Wyner's original scheme [1] where only one such recipe needed to be memorized. Therefore, in the remainder of the proof, we can and shall treat the process of distilling a secret key as if Eve's channel were known to all parties in advance, without losing any generality. That is, for the sake of using familiar notation and terminology, we shall talk about "encoding" a secret key at Alice, for "transmission" to Bob, although the secret key is only agreed upon at the end of the frame.

Second, a secret key $\mathbf{K} = \bigcup_{j \in \mathscr{I}_{ne}} \mathbf{K}_j$ is "encoded" into all encoding levels $j$ belonging to $\mathscr{I}_{ne}$, i.e. over levels belonging to both $\mathscr{I}_k$ and $\mathscr{I}_n$. Recall that Bob cannot decode the levels of $\mathscr{I}_n$. We do this because it is easier to prove that the whole key $\mathbf{K}$ is secret to Eve. Once this is accomplished, we shall follow a simple argument of [15] (which we replicate in (33) for completeness) to show that the sub-key $\bigcup_{j \in \mathscr{I}_k} \mathbf{K}_j$, which can actually be decoded and used by Bob, is also perfectly secret.

We use a separate secret key encoding for each of Alice's encoding levels in $\mathscr{I}_{ne}$. As a consequence, Eve sees a fast fading multiple access channel, where the transmitters have different power constraints, but the same channel coefficient. In this context, we note that the conditions set forth for the rates $R_{j,E}$ in (20) and (21) are exactly the conditions necessary for these rates to belong to the boundary of the capacity region of Eve's equivalent multiple access channel. The problem of a multiple access eavesdropper AWGN channel was discussed in [15]. However, some of the results in [15] do not necessarily reflect our views. We provide an explanation of this assertion in Appendix B. Therefore, we continue with describing a correct encoding method which yields an achievable secret key rate.

For any level of encoding $j \in \mathscr{I}_{ne}$, we encode a secret key $\mathbf{K}_j$ according to Wyner's scheme [1], [17]. That is, if $j \neq 1$, we randomly bin the randomly generated $N$-dimensional codebook of $2^{NR_j}$ codewords into $2^{N(R_j - R_{E,j})}$ bins. The secret message corresponds to the index of the bin, while the exact codeword in the bin is randomly picked. The rates $R_{E,j}$ are selected as in the statement of the theorem. If $j = 1 \in \mathscr{I}_{ne}$ (recall that codebook 1 was already binned once), Bob generates the bins in two steps: first he identifies the $2^{N(R_1 - R_s)}$ bins used for transmitting Alice's encrypted message, and then he randomly *groups* these bins into $2^{N(R_1 - R_{E,1})}$ larger bins. A secret message is encoded into the indices of the resulting larger bins.

Denote the resulting $N$-dimensional output sequence of level $j$ by $\mathbf{X}_j$, and denote the $p$-th component of $\mathbf{X}_j$ by $\mathbf{X}_j(p)$. Also denote the union of the $N$-sequences from all levels (including those from $\mathscr{I}_e$ which do not carry a secret key) by $\mathbf{X} = \bigcup_{j \in \hat{\mathscr{I}}} \mathbf{X}_j$. The notation $\mathbf{X}(p)$ now denotes the $n$-dimensional set consisting of the $p$-th components of the output sequences from every encoding level, that is $\mathbf{X}(p) = \bigcup_{j \in \hat{\mathscr{I}}} \mathbf{X}_j(p)$. Eve's received sequence is now $\mathbf{Z} = \mathbf{H}_W \cdot \sum_{j \in \hat{\mathscr{I}}} \mathbf{X}_j + \mathbf{Q}$, where $\mathbf{H}_W$ is the $N$-dimensional vector of channel realizations corresponding to the $N$ symbols, $\mathbf{Q}$ is Eve's $N$-dimensional additive white Gaussian noise sequence, and $(\cdot)$ denotes component-wise multiplication. The $p$-th scalar components of these vectors will be denoted by $\mathbf{Z}(p)$, $\mathbf{H}_W(p)$ and $\mathbf{Q}(p)$, respectively. The notation $\mathbf{X}_{\mathscr{S}}$ will be used for the union of the output sequences corresponding to levels with indices in $\mathscr{S}$, i.e. $\mathbf{X}_{\mathscr{S}} = \bigcup_{j \in \mathscr{S}} \mathbf{X}_j$, and the notation for the $p$-th components is extended correspondingly.

Eve's equivocation about the secret key can be written as follows

$$
\begin{aligned}
\Delta &= \frac{H(\mathbf{K}|\mathbf{Z}, \mathbf{H}_W)}{H(\mathbf{K})} = \frac{H(\mathbf{K}, \mathbf{Z}, \mathbf{H}_W) - H(\mathbf{Z}, \mathbf{H}_W)}{H(\mathbf{K})} \overset{(a)}{=} \\
&= \frac{H(\mathbf{K}) + H(\mathbf{Z}, \mathbf{H}_W, \mathbf{X}|\mathbf{K})}{H(\mathbf{K})} - \\
&\quad - \frac{H(\mathbf{X}|\mathbf{Z}, \mathbf{H}_W, \mathbf{K}) + H(\mathbf{Z}, \mathbf{H}_W)}{H(\mathbf{K})} \overset{(b)}{=} \\
&= \frac{H(\mathbf{K}) + H(\mathbf{Z}, \mathbf{H}_W|\mathbf{X}, \mathbf{K}) + H(\mathbf{X}|\mathbf{K})}{H(\mathbf{K})} - \\
&\quad - \frac{H(\mathbf{X}|\mathbf{Z}, \mathbf{H}_W, \mathbf{K}) + H(\mathbf{Z}, \mathbf{H}_W)}{H(\mathbf{K})} \overset{(c)}{=} \\
&= 1 - \frac{I(\mathbf{X}; \mathbf{Z}, \mathbf{H}_W) - I(\mathbf{X}; \mathbf{Z}, \mathbf{H}_W|\mathbf{K})}{H(\mathbf{K})}, \quad (22)
\end{aligned}
$$

where both $(a)$ and $(b)$ result from the chain rule for entropy, while (c) from the fact that $\mathbf{K} \to \mathbf{X} \to \mathbf{Z}$ form a Markov chain.

Denote $\mathcal{D} = I(\mathbf{X}; \mathbf{Z}, \mathbf{H}_W) - I(\mathbf{X}; \mathbf{Z}, \mathbf{H}_W|\mathbf{K})$. We can now write

$$
\begin{aligned}
I(\mathbf{X}; \mathbf{Z}, \mathbf{H}_W) = H(\mathbf{X}_{\mathscr{I}_e}) + H(\mathbf{X}_{\mathscr{I}_{ne}}) - \\
- H(\mathbf{X}_{\mathscr{I}_e}|\mathbf{Z}, \mathbf{H}_W) - H(\mathbf{X}_{\mathscr{I}_{ne}}|\mathbf{X}_{\mathscr{I}_e}, \mathbf{Z}, \mathbf{H}_W),
\end{aligned} \quad (23)
$$

$$
H(\mathbf{X}|\mathbf{K}) = H(\mathbf{X}_{\mathscr{I}_e}) + H(\mathbf{X}_{\mathscr{I}_{ne}}|\mathbf{K}), \quad (24)
$$

and

$$
\begin{aligned}
H(\mathbf{X}|\mathbf{Z}, \mathbf{H}_W, \mathbf{K}) = \\
= H(\mathbf{X}_{\mathscr{I}_e}|\mathbf{Z}, \mathbf{H}_W, \mathbf{K}) + H(\mathbf{X}_{\mathscr{I}_{ne}}|\mathbf{X}_{\mathscr{I}_e}, \mathbf{Z}, \mathbf{H}_W, \mathbf{K}) \leq \\
\leq H(\mathbf{X}_{\mathscr{I}_e}|\mathbf{Z}, \mathbf{H}_W) + H(\mathbf{X}_{\mathscr{I}_{ne}}|\mathbf{X}_{\mathscr{I}_e}, \mathbf{Z}, \mathbf{H}_W, \mathbf{K}),
\end{aligned} \quad (25)
$$

where we used the fact that $\{\mathbf{X}_j : j \in \mathscr{I}\}$ are all independent of each other, and that conditioning reduces entropy. Substituting (23)-(25) in the expression of $\mathcal{D}$ above, and noting that $H(\mathbf{X}_{\mathscr{I}_{ne}}) = H(\mathbf{X}_{\mathscr{I}_{ne}}|\mathbf{X}_{\mathscr{I}_e})$, we obtain

$$
\begin{aligned}
\mathcal{D} \leq I(\mathbf{X}_{\mathscr{I}_{ne}}; \mathbf{Z}, \mathbf{H}_W|\mathbf{X}_{\mathscr{I}_e}) - H(\mathbf{X}_{\mathscr{I}_{ne}}|\mathbf{K}) + \\
+ H(\mathbf{X}_{\mathscr{I}_{ne}}|\mathbf{X}_{\mathscr{I}_e}, \mathbf{Z}, \mathbf{H}_W, \mathbf{K}).
\end{aligned} \quad (26)
$$

By the code construction, and recalling that the rates $R_{E,j}$ in the statement of the theorem are picked such that they belong to the boundary of the capacity region of Eve's equivalent multiple access channel (they satisfy (21)), we can use Fano's inequality, the union bound and arguments similar to those used in driving equation (78) of [1], to upper bound

$$
H(\mathbf{X}_{\mathscr{I}_{ne}}|\mathbf{X}_{\mathscr{I}_e}, \mathbf{Z}, \mathbf{H}_W, \mathbf{K}) \leq N \delta_N, \quad (27)
$$

where $\delta_N \to 0$ as $N \to \infty$. This is quite intuitive, since given the secret key, the other information is transmitted by Alice using codes which are good for Eve's multiple access channel. In fact

$$
\delta_N = \sum_{j \in \mathscr{I}_{ne}} \frac{1}{N} h(p_{e,j}) + p_{e,j} R_{E,j}, \quad (28)
$$

where $p_{e,j}$ is the probability of error for the layer-$j$ such code, and $h(\cdot)$ is the binary entropy function $h(x) = -x \log_2(x) -$ $(1-x)\log_2(1-x)$. Since the random, complementary-to-the-secret-key information is carried by these codes at a total rate almost equal to the capacity of the virtual MAC between Alice and Eve, corresponding to the encoding levels in $\mathscr{I}_{ne}$, we also have

$$H(\mathbf{X}_{\mathscr{I}_{ne}}|\mathbf{K}) = Nq\mathbf{E}_{h_W}\left[\log\left(1+\right.\right.$$
$$\left.\left.+\frac{\sum_{j\in\mathscr{I}_{ne}}(1-\alpha_j)\alpha_{j-1}\ldots\alpha_1\mathscr{P}h_W}{\sigma_N^2}\right)\right] - N\epsilon. \quad (29)$$

To upper bound the first term on the right hand side of (26), we write

$$I(\mathbf{X}_{\mathscr{I}_{ne}};\mathbf{Z},\mathbf{H}_W|\mathbf{X}_{\mathscr{I}_e}) =$$
$$= H(\mathbf{Z},\mathbf{H}_W|\mathbf{X}_{\mathscr{I}_e}) - H(\mathbf{Z},\mathbf{H}_W|\mathbf{X}_{\mathscr{I}}) \stackrel{(a)}{=}$$
$$= H(\mathbf{Z},\mathbf{H}_W|\mathbf{X}_{\mathscr{I}_e}) - NH(\mathbf{Z}(p),\mathbf{H}_W(p)|\mathbf{X}_{\mathscr{I}}(p)) \stackrel{(b)}{\leq}$$
$$\leq NH(\mathbf{Z}(p),\mathbf{H}_W(p)|\mathbf{X}_{\mathscr{I}_e}(p)) -$$
$$-NH(\mathbf{Z}(p),\mathbf{H}_W(p)|\mathbf{X}_{\mathscr{I}}(p)) =$$
$$= NI(\mathbf{X}_{\mathscr{I}_{ne}}(p);\mathbf{Z}(p),\mathbf{H}_W(p)|\mathbf{X}_{\mathscr{I}_e}(p)) \stackrel{(c)}{\leq}$$
$$\leq Nq\mathbf{E}_{h_W}\left[\log\left(1+\right.\right.$$
$$\left.\left.+\frac{\sum_{j\in\mathscr{I}_{ne}}(1-\alpha_j)\alpha_{j-1}\ldots\alpha_1\mathscr{P}h_W}{\sigma_N^2}\right)\right]. \quad (30)$$

Equality in $(a)$ follows from the fact that the channel is memoryless, $(b)$ follows from the chain rule for entropy and the fact that conditioning does not increase entropy, and $(c)$ is obtained by using Jensen's inequality, as in the proof of the converse to the AWGN channel coding theorem in Section 9.2. of [18].

Putting together (27), (29) and (30), we obtain

$$\mathcal{D} \leq N(\epsilon+\delta_N), \quad (31)$$

which in turn implies

$$\Delta \geq 1 - N\frac{\epsilon+\delta_N}{H(\mathbf{K})}. \quad (32)$$

Since $\frac{H(\mathbf{K})}{N}$ is a constant, the right-hand side of (32) converges to 1 as $N \to \infty$. Thus, we have proved that the key $\mathbf{K}$ remains secret from Eve as long as the codeword length $N$ goes to infinity. However, note that the entire key $\mathbf{K}$ cannot be understood by Bob. In fact, Bob and Alice can only agree on the part $\mathbf{K}_{\mathscr{I}_k}$ of the key. But the secrecy of the entire key guarantees the secrecy of any part of the key [15]. For the sake of completeness, we restate the following proof from [15].

$$H(\mathbf{K}_{\mathscr{I}_k}|\mathbf{Z},\mathbf{H}_W) \stackrel{(a)}{=}$$
$$= H(\mathbf{K}_{\mathscr{I}_{ne}}|\mathbf{Z},\mathbf{H}_W) - H(\mathbf{K}_{\mathscr{I}_n}|\mathbf{K}_{\mathscr{I}_k},\mathbf{Z},\mathbf{H}_W) \stackrel{(b)}{\geq}$$
$$\geq H(\mathbf{K}) - N(\epsilon+\delta_N) -$$
$$-H(\mathbf{K}_{\mathscr{I}_n}|\mathbf{K}_{\mathscr{I}_k},\mathbf{Z},\mathbf{H}_W) \stackrel{(c)}{\geq}$$
$$\geq H(\mathbf{K}_{\mathscr{I}_k}) + H(\mathbf{K}_{\mathscr{I}_n}) - N(\epsilon+\delta_N) -$$
$$-H(\mathbf{K}_{\mathscr{I}_n}|\mathbf{K}_{\mathscr{I}_k},\mathbf{Z},\mathbf{H}_W) \stackrel{(d)}{\geq}$$
$$\geq H(\mathbf{K}_{\mathscr{I}_k}) - N(\epsilon+\delta_N), \quad (33)$$

where $(a)$ follows from the chain rule, $(b)$ from (32) and the definition of $\Delta$, $(c)$ from the independence of the keys from different encoding levels, and $(d)$ from the fact that conditioning does not increase entropy.

This results in

$$\frac{H(\mathbf{K}_{\mathscr{I}_k}|\mathbf{Z},\mathbf{H}_W)}{H(\mathbf{K}_{\mathscr{I}_k})} \geq 1 - N\frac{\epsilon+\delta_N}{H(\mathbf{K}_{\mathscr{I}_k})} \to 1 \quad (34)$$

as $N \to \infty$ (because $\frac{H(\mathbf{K}_{\mathscr{I}_k})}{N}$ is a constant). Therefore, a perfectly secret key can be distilled from the random information transmitted by the encoding levels in $\mathscr{I}_k$. ∎

We have seen the best achievable secret key rate if $q \in [q_{i-1}, q_i)$. The next theorem provides Eve's optimal strategy, i.e. which is the most destructive value of $q$ under the present conservative scenario, and also Alice's best achievable secrecy rate under this eavesdropper strategy. Note that although the game between Eve and the legitimate parties Alice and Bob is a dynamic one, where Bob and Alice need to re-evaluate their strategies over each frame, the following result can be thought of as some form of equilibrium. Indeed, under our assumptions, it is sub-optimal for either Eve or the Alice/Bob pair to deviate from the following strategies.

*Theorem 5:* Consider a given quantization $\{q_i|i \in \widehat{\mathscr{I}}\}$ of the interval $[0,1]$, and a given power allocation between the corresponding encoding levels $\{\alpha_i|i \in \widehat{\mathscr{I}}\}$.

(1) If Eve's optimal value of $q$ is such that $q \in [q_{i-1}, q_i)$, then $q$ is arbitrarily close to $q_i$.

(2) Eve's optimal strategy $q$ under our conservative scenario is the same over all frames.

(3) Denote the achievable secret key rates by $\{R_{k,i} : i = 1, 2, \ldots, n\}$, where $R_{k,i}$ is the best achievable secret key rate given by Theorem 4, under $q = q_i$. Then Eve's optimal strategy is $q_{i_{opt}} = \arg\min_{q_i}\{R_{k,i}\}$, if $\min_{q_i}\{R_{k,i}\} < 0.5R_1$, and $q_{i_{opt}} = q_1$, otherwise.

(4) Under Eve's optimal strategy, the maximum achievable secrecy rate (under the current setup) is

$$R_s = \min\{0.5R_1, R_{k,i_{opt}}\}. \quad (35)$$

(5) There is no loss of performance incurred by restricting the rate of the encrypted message to $R_s \leq 0.5R_1$ in (15).

*Proof:* (1) Using Theorem 4, it is easy to check that, given $q \in [q_{i-1}, q_i)$, the achievable secret key rate is a decreasing function of $q$. Therefore, if $q \in [q_{i-1}, q_i)$, Eve's optimal strategy is to pick $q$ arbitrarily close to $q_i$.

(2),(3),(4) We have already mentioned that our encoding strategy restricts the rate of the encrypted message to $R_s \leq 0.5R_1$, by restricting the rate of generation of the secret key to $R_k \leq 0.5R_1$. If $\min_{q_i}\{R_{k,i}\}$ is achieved by $q_{i_{opt}}$ and is less than $0.5R_1$, then by switching to a different strategy $q_d$, Eve will only increase the rate of generation of the secret key, and hence the rate of transmission of the encrypted message. On the other hand, if $\min_{q_i}\{R_{k,i}\} \geq 0.5R_1$, then no matter what Eve's strategy is, the secrecy rate will be constrained by the encoding scheme to $0.5R_1$.

(5) The constraint introduced by the encoding scheme is a conservative one. Although the secret key may be generated by multiple layers, at the end of a frame, the encrypted message is transmitted only by the first encoding level. This is because neither Alice, nor Bob know the channel quality in advance, and thus, to ensure reliable decoding of the secret message, they have to plan for the worse. For example, Eve might choose to constantly play a strategy $q \in [0, q_1)$, which implies that Bob will only be able to decode level 1 of the code. This message, transmitted at a maximum rate of $R_1$, has to carry the encrypted message and generate a secret key, simultaneously. But since Eve's strategy remains in $[0, q_1)$ over the next frames, the rate of the encrypted message cannot exceed $0.5 R_1$ – there would not be enough secret key bits to encrypt it. Therefore, the strategy $q \in [0, q_1)$ can function as a "default" state for Eve, where she could take refuge if the achievable secrecy rate under any other strategy exceeded $0.5 R_1$. ∎

Theorems 4 and 5 above offer a good description of the achievable secrecy rates. However, in Theorem 4 we assumed that the set $\mathscr{I}_e$ of indices corresponding to the levels that are *perfectly decodable by Eve*, and the sets $\mathscr{I}_k$ and $\mathscr{I}_n$ are readily available. Due to the fact that our encoding scheme is designed such that Bob should perform successive interference cancellation, the set of levels that are not decodable by Bob is easy to compute. However, the characterization of the set $\mathscr{I}_e$ and its complement $\mathscr{I}_{ne}$ is not straightforward. The following proposition shows how these sets can be found. Its proof results from Lemma 8 in Appendix C.

*Proposition 6:* The maximal set of indices $\mathscr{I}_e$ corresponding to the levels that are perfectly decodable by Eve is the largest of the sets $\mathscr{V}_e$ for which

$$\sum_{j \in \mathscr{S}} R_j \leq q \mathbf{E}_{h_W} \bigg[ \log \bigg( 1 + \frac{\sum_{j \in \mathscr{S}} (1-\alpha_j) \alpha_{j-1} \ldots \alpha_1 \mathscr{P} h_W}{\sigma_N^2 + \sum_{i \in \mathscr{V}_e^c} (1-\alpha_i) \alpha_{i-1} \ldots \alpha_1 \mathscr{P} h_W} \bigg) \bigg], \ \forall \mathscr{S} \subseteq \mathscr{V}_e, \quad (36)$$

where $\mathscr{V}_e^c$ is the complement of $\mathscr{V}_e$ with respect to $\widehat{\mathscr{I}}$

### C. On the Complexity of the Algorithm: Selecting $\{\alpha_i\}$ and $\{q_i\}$.

Our results so far facilitate the computation of an achievable secrecy rate, given a partition of the interval $[0,1]$ expressed in terms of the parameters $\{q_1, q_2, \ldots, q_{n-1}\}$, and a power allocation between the encoding levels, given by the parameters $\{\alpha_1, \alpha_2, \ldots, \alpha_{n-1}\}$. If Alice and Bob wish to exploit the full secrecy capabilities of the model, they should perform a maximization of the achievable secrecy rate with respect to the parameters $\{(q_i, \alpha_i) : i = 1, 2, \ldots, n-1\}$.

The optimization problem requires a high complexity numerical algorithm. Recall that for each value of the parameter vector $\{(q_i, \alpha_i) : i = 1, 2, \ldots, n-1\}$ we need to find the set $\mathscr{I}_e$ as in the Proposition 6, which involves combinatorial complexity. Given $\mathscr{I}_e$, we need to find the optimal set of encoding layers that should be involved in the elaboration of the secret key. The following example illustrates the steps of the algorithm for the least involved scenario: that of $n = 2$.

**A simple case:** $n = 2$

We start by selecting a value for the parameter vector $(q_1, \alpha_1)$. The transmission rates for the two encoding levels

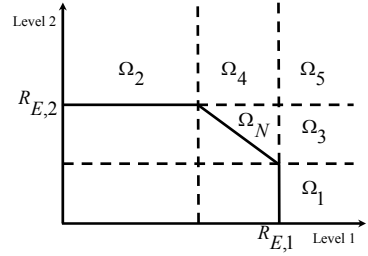

Fig. 5. Eve's equivalent-MAC capacity region.

become:

$$R_1 = \mathbf{E}_{h_M} \bigg[ \log \bigg( 1 + \frac{(1-\alpha_1)\mathscr{P} h_M}{\sigma_N^2 + \alpha_1 \mathscr{P} h_M + \mathscr{J}} \bigg) \bigg], \quad (37)$$

and

$$R_2 = \mathbf{E}_{h_M} \bigg[ q_1 \log \bigg( 1 + \frac{\alpha_1 \mathscr{P} h_M}{\sigma_N^2} \bigg) + \\ + (1-q_1) \log \bigg( 1 + \frac{\alpha_1 \mathscr{P} h_M}{\sigma_N^2 + \frac{\mathscr{J}}{1-q_1}} \bigg) \bigg]. \quad (38)$$

To illustrate all possible cases, we shall refer to Figure 5, where we represented the (equivalent-) MAC capacity region of Eve. Although this region depends on Eve's strategy $q \in \{q_1, 1\}$, we shall use it as a reference frame, in which the tuple $(R_1, R_2)$ (which is fixed as above) can occupy different positions, depending on $q$.

Depending on Eve's strategy $q \in \{q_1, 1\}$, we have the following two algorithms:

If $q = q_1$, only level 1 is intelligible to Bob.

1) If

$$q_1 \mathbf{E}_{h_W} \bigg[ \log \bigg( 1 + \frac{(1-\alpha_1)\mathscr{P} h_W}{\sigma_N^2} \bigg) \bigg] \geq R_1, \quad (39)$$

$$q_1 \mathbf{E}_{h_W} \bigg[ \log \bigg( 1 + \frac{\alpha_1 \mathscr{P} h_W}{\sigma_N^2} \bigg) \bigg] \geq R_2, \quad (40)$$

$$q_1 \mathbf{E}_{h_W} \bigg[ \log \bigg( 1 + \frac{\mathscr{P} h_W}{\sigma_N^2} \bigg) \bigg] \geq R_1 + R_2, \quad (41)$$

or, equivalently, if the tuple $(R_1, R_2)$ is inside the capacity region of Figure 5, then $\mathscr{I}_e = \{1, 2\}$ (i.e. both levels are perfectly decodable by Eve.) In this case, no secret key may be generated, and no secret message may be transmitted.

2) Else, if

$$q_1 \mathbf{E}_{h_W} \bigg[ \log \bigg( 1 + \frac{(1-\alpha_1)\mathscr{P} h_W}{\sigma_N^2 + \alpha_1 \mathscr{P} h_W} \bigg) \bigg] \geq R_1, \quad (42)$$

which, along with the condition that we are outside the capacity region implies

$$q_1 \mathbf{E}_{h_W} \bigg[ \log \bigg( 1 + \frac{\alpha_1 \mathscr{P} h_W}{\sigma_N^2} \bigg) \bigg] < R_2, \quad (43)$$

that is $(R_1, R_2)$ is in the region $\Omega_2$ of Figure 5, then $\mathscr{I}_e = \{1\}$. But since level 2 is not intelligible to Bob, no secret key may be generated.

3) Else, if

$$q_1 \mathbf{E}_{h_W}\left[\log\left(1 + \frac{\alpha_1 \mathscr{P} h_W}{\sigma_N^2 + (1-\alpha_1)\mathscr{P} h_W}\right)\right] \geq$$
$$\geq R_2, \quad (44)$$

which (since not in the capacity region) implies that $(R_1, R_2)$ is in $\Omega_1$, i.e.

$$q_1 \mathbf{E}_{h_W}\left[\log\left(1 + \frac{(1-\alpha_1)\mathscr{P} h_W}{\sigma_N^2}\right)\right] < R_1, \quad (45)$$

then $\mathscr{I}_e = \{2\}$. For this scenario, only level 1 may generate a secret key at a rate equal to $\min\{0.5R_1, R_1 - q_1\mathbf{E}_{h_W}\left[\log\left(1 + \frac{(1-\alpha_1)\mathscr{P} h_W}{\sigma_N^2}\right)\right]\}$.

4) The remaining case is when neither of the two encoding levels is intelligible to Eve. Under this assumption, both levels 1 and 2 may be involved in the generation of the secret key. A secret key may be generated at rate $\min\{0.5R_1, R_1 - R_{E,1}\}$, where $R_{E,1}$ is either equal to $q_1 \mathbf{E}_{h_W}\left[\log\left(1 + \frac{(1-\alpha_1)\mathscr{P} h_W}{\alpha_1 \mathscr{P} h_W + \sigma_N^2}\right)\right]$ if $(R_1, R_2)$ is in $\Omega_4 \bigcup \Omega_5$ of Figure 5, or $R_{E,1} = q_1 \mathbf{E}_{h_W}\left[\log\left(1 + \frac{\mathscr{P} h_W}{\sigma_N^2}\right)\right]\} - R_2$ if $(R_1, R_2)$ is in $\Omega_3 \bigcup \Omega_N$.

If $q = 1$, both levels are intelligible to Bob.

1) If $(R_1, R_2)$ is in Eve's capacity region, i.e.

$$\mathbf{E}_{h_W}\left[\log\left(1 + \frac{(1-\alpha_1)\mathscr{P} h_W}{\sigma_N^2}\right)\right] \geq R_1, \quad (46)$$

$$\mathbf{E}_{h_W}\left[\log\left(1 + \frac{\alpha_1 \mathscr{P} h_W}{\sigma_N^2}\right)\right] \geq R_2, \quad (47)$$

$$\mathbf{E}_{h_W}\left[\log\left(1 + \frac{\mathscr{P} h_W}{\sigma_N^2}\right)\right] \geq R_1 + R_2, \quad (48)$$

then $\mathscr{I}_e = \{1, 2\}$ (i.e. both levels are perfectly decodable by Eve), and no secret key may be generated.

2) Else, if $(R_1, R_2) \in \Omega_2$, i.e.

$$\mathbf{E}_{h_W}\left[\log\left(1 + \frac{(1-\alpha_1)\mathscr{P} h_W}{\sigma_N^2 + \alpha_1 \mathscr{P} h_W}\right)\right] \geq R_1, \quad (49)$$

and

$$\mathbf{E}_{h_W}\left[\log\left(1 + \frac{\alpha_1 \mathscr{P} h_W}{\sigma_N^2}\right)\right] < R_2, \quad (50)$$

then $\mathscr{I}_e = \{1\}$. For this scenario, level 2 may generate a secret key at rate $R_2 - \mathbf{E}_{h_W}\left[\log\left(1 + \frac{\alpha_1 \mathscr{P} h_W}{\sigma_N^2}\right)\right]$.

3) Else, if $(R_1, R_2) \in \Omega_1$, i.e.

$$\mathbf{E}_{h_W}\left[\log\left(1 + \frac{\alpha_1 \mathscr{P} h_W}{\sigma_N^2 + (1-\alpha_1)\mathscr{P} h_W}\right)\right] \geq R_2, \quad (51)$$

and

$$\mathbf{E}_{h_W}\left[\log\left(1 + \frac{(1-\alpha_1)\mathscr{P} h_W}{\sigma_N^2}\right)\right] < R_1, \quad (52)$$

then $\mathscr{I}_e = \{2\}$. For this scenario, level 1 may generate a secret key at rate $\min\{0.5R_1, R_1 - \mathbf{E}_{h_W}\left[\log\left(1 + \frac{(1-\alpha_1)\mathscr{P} h_W}{\sigma_N^2}\right)\right]\}$.

4) When neither of the two encoding levels is intelligible to Eve, both levels may be involved in the generation of the secret key. We can achieve the secret key rate given by $R_1 + R_2 - R_{E,1} - R_{E,2}$, where $R_{E,1}$ and $R_{E,2}$ are chosen such that $R_{E,1} + R_{E,2} = \mathbf{E}_{h_W}\left[\log\left(1 + \frac{\mathscr{P} h_W}{\sigma_N^2}\right)\right]$, $0.5R_1 \leq R_{E,1} \leq \min\{R_1, \mathbf{E}_{h_W}\left[\log\left(1 + \frac{(1-\alpha_1)\mathscr{P} h_W}{\sigma_N^2}\right)\right]\}$, and $R_{E,2} \leq \min\{R_2, \mathbf{E}_{h_W}\left[\log\left(1 + \frac{\alpha_1 \mathscr{P} h_W}{\sigma_N^2}\right)\right]\}$, with the observations following Theorem 4.

Since we are currently investigating the conservative scenario, Eve will pick the strategy ($q = q_1$ or $q = 1$) which yields the minimum secrecy rate. We have to find the value of $(q_1, \alpha_1)$ which yields the largest such minimum. Equivalently, the optimal $(q_1, \alpha_1)$ will provide equal achievable secrecy rates for $q = q_1$ and for $q = 1$.

It is important to note that, although the algorithm may be extremely complex, it needs to be solved only once for the desired value of $n$. The optimal parameters may then be stored at both legitimate parties.

In an effort to reduce the complexity of the algorithm, we propose to pick the parameters $\{(q_i) : i = 1, \ldots, n-1\}$ such that $\{q_0, q_1, q_2, \ldots, q_{n-1}, q_n\}$ are all equally spaced, which corresponds to a uniform partition (or "quantization") of the interval $[0, 1]$. With this rule in place, the optimization needs to be performed only over the $(n-1)$ parameters $\alpha_1, \ldots, \alpha_{n-1}$, hence the complexity is reduced by half.

From our numerical results for $n = 2$ and $n = 3$ (see Figure 8), the loss of optimality due to the uniform partition of $[0, 1]$ is not very significant. We conjecture that, as $n$ increases, this loss of performance should become negligible. Our remark is based on the fact that as $n \to \infty$ the optimal partition of the interval $[0, 1]$ approaches a uniform partition (with a vanishing step).

IV. NUMERICAL RESULTS

In Figures 6 and 7 we show the improvement of our BMW secrecy encoding scheme over the worst-case scenario approach of (10). Note that if Eve's channel coefficient is close (statistically) to Bob's – the case of Figure 6 – the worst-case approach of (10) – or equivalently the case $n = 1$ – cannot achieve a positive secrecy rate.

However, even Wyner's pure scheme implemented as in (10) can achieve a positive secrecy rate if $\lambda_W > \lambda_M(1 + \frac{\mathscr{P}}{\sigma_N^2})$, as discussed in Section II – see Figure 7. The merit of our novel encoding scheme is significant.

The best-case (or "minimax") scenario solution of [12] is given in both Figures 6 and 7 for comparison. The "minimax" scenario describes a situation when Alice and Bob can know Eve's strategy in advance (or, in game-theoretic terms, Alice *plays first*). Although this scenario may not seem like a reasonable model, it serves as an upper-bound on the achievable secrecy rates.

Figure 8 depicts the performance of the BMW secrecy encoding scheme when the partition of the interval $[0, 1]$ into intervals of the form $[q_{i-1}, q_1)$ is done uniformly, i.e. the parameters $q_0, q_1, q_2, \ldots, q_n$ are equally spaced, instead of being picked in an optimal way. We note that the degradation of the achievable secrecy rates is quite small and decreasing as $n$ increases. Figures 9 and 10 show the design parameters used for obtaining the results of Figure 8.

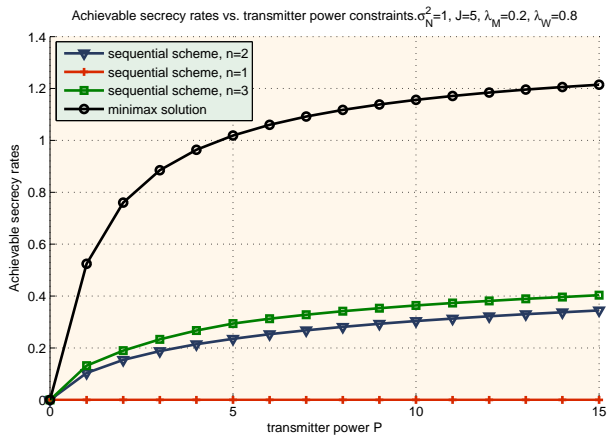

Fig. 6. Achievable secrecy rates with our BMW secrecy encoding scheme. Exponentially distributed channel coefficients with $\lambda_M = 0.3$, $\lambda_W = 0.8$, $J = 5$, $\sigma_N^2 = 1$.

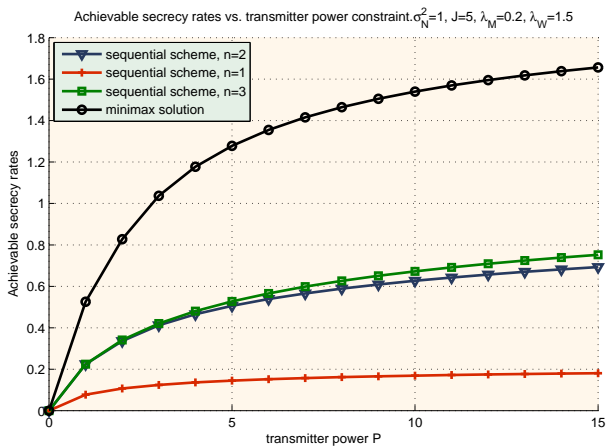

Fig. 7. Achievable secrecy rates with our BMW secrecy encoding scheme. Exponentially distributed channel coefficients with $\lambda_M = 0.2$, $\lambda_W = 1.5$, $J = 5$, $\sigma_N^2 = 1$.

## V. CONCLUSIONS

We have seen how an active eavesdropper can seriously decrease the achievable secrecy rate in a classical scenario of a fast-fading AWGN channel with an eavesdropper. Our scenario models the most conservative and most practical approach to the active eavesdropper.

We have seen that, in order to take advantage of the non-duplex nature of the eavesdropper's terminal, we need a more elaborate, block-Markov Wyner encoding scheme. While in the classical eavesdropper scenario the legitimate receiver is completely passive, our scheme relies heavily on the cooperation of the receiver. That means that at the end of each frame, Bob is required to feed back to Alice information about Eve's strategy, and then, based on this information, replicate Alice's efforts to distill a secret key.

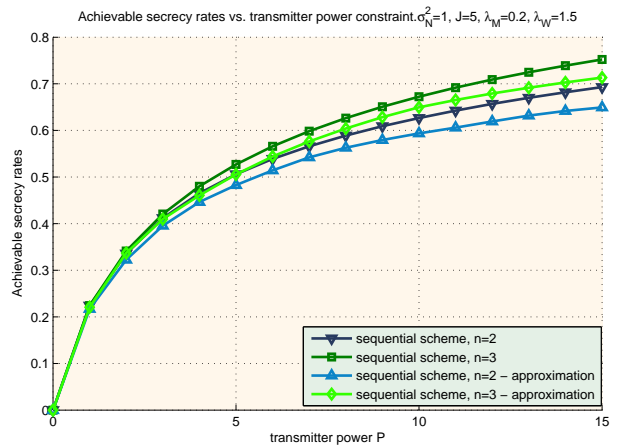

Fig. 8. Achievable secrecy rates with our BMW secrecy encoding scheme, with uniform and with optimized partition of the interval $[0, 1]$. Exponentially distributed channel coefficients with $\lambda_M = 0.2$, $\lambda_W = 1.5$, $J = 5$, $\sigma_N^2 = 1$.

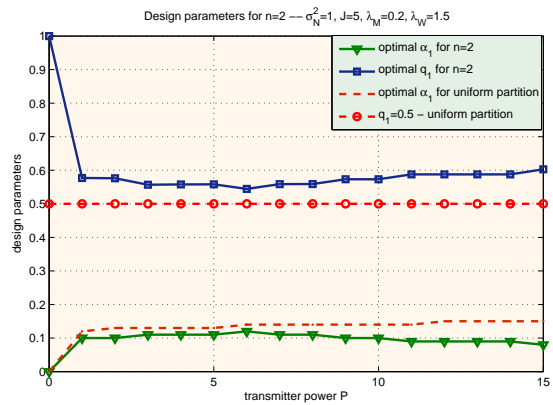

Fig. 9. The encoding parameters $q_1$ and $\alpha_1$ for the case $n = 2$: optimal and uniform partition of the interval [0,1]. Exponentially distributed channel coefficients with $\lambda_M = 0.2$, $\lambda_W = 1.5$, $J = 5$, $\sigma_N^2 = 1$.

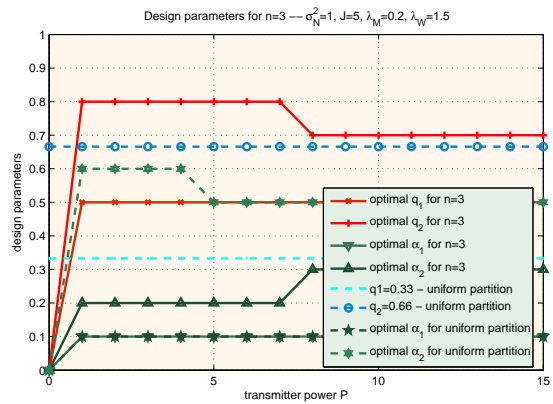

Fig. 10. The encoding parameters $q_1$, $q_2$, $\alpha_1$ and $\alpha_2$ for the case $n = 3$: optimal and uniform partition of the interval [0,1]. Exponentially distributed channel coefficients with $\lambda_M = 0.2$, $\lambda_W = 1.5$, $J = 5$, $\sigma_N^2 = 1$.

Although the performance of our BMW scheme remains below the secrecy rate upper-bound provided by the best-case scenario of [12], the improvement it brings over the passive-receiver solution is quite significant.

## APPENDIX A
## A USEFUL LEMMA

The following lemma is used several times in this paper.
*Lemma 7:* The function

$$f(q) = q \log(1+x) + (1-q) \log \left(1 + \frac{x}{1+\frac{y}{1-q}}\right), \quad (53)$$

where $x, y > 0$, is strictly increasing and strictly convex as a function of $q$.

*Proof:* It is straightforward to compute

$$\frac{df(q)}{dq} = \log \left((1+x) \frac{1+\frac{y}{1-q}}{1+x+\frac{y}{1-q}}\right) - \\ - \frac{xy}{1-q} \cdot \frac{1}{(1+\frac{y}{1-q})(1+x+\frac{y}{1-q})}, \quad (54)$$

and

$$\frac{d^2 f(q)}{dq^2} = \frac{\frac{xy}{(1-q)^2}}{(1+\frac{y}{1-q})(1+x+\frac{y}{1-q})} \cdot \\ \cdot \left[1 - \frac{1+x-(\frac{y}{(1-q)})^2}{(1+\frac{y}{1-q})(1+x+\frac{y}{1-q})}\right]. \quad (55)$$

Since $1 + \frac{y}{1-q} > 1$ and $1 + x - (\frac{y}{(1-q)})^2 < 1 + x + \frac{y}{1-q}$, we can state that $\frac{d^2 f(q)}{dq^2} > 0$. Therefore, $\frac{df(q)}{dq}$ is a strictly increasing function of $q$. But evaluating the first derivative in $q = 0$ we get

$$\frac{df}{dq}(0) = \\ = \log \left(\frac{(1+x)(1+y)}{1+x+y}\right) - \frac{xy}{(1+y)(1+x+y)} = \\ = \log \left(1 + \frac{xy}{1+x+y}\right) - \frac{xy}{(1+x+y)(1+y)} \overset{(a)}{\geq} \\ \geq \frac{xy}{(1+x)(1+y)} - \frac{xy}{(1+x+y)(1+y)} \overset{(b)}{>} 0, \quad (56)$$

where inequality $(a)$ follows from $\log(1+\beta) > \frac{\beta}{1+\beta}$ for any $\beta > -1$, $\beta \neq 0$, if we replace $\beta = \frac{xy}{1+x+y}$, while inequality $(b)$ follows since $x > 0$. Therefore $\frac{df(q)}{dq}$ is always strictly positive and strictly increasing, which implies that $f(q)$ is strictly increasing and strictly convex. ∎

## APPENDIX B
## COMMENTS ON PREVIOUSLY EXISTING RESULTS ON MAC SECRECY

The most notable recent results on the achievable secrecy rates for MACs are provided by [15] and [16]. Although the papers bring unquestionable contributions relevant to our own scenario, such as the concepts of individual and collective secrecy, and the improvement of the secrecy sum-rates by noise injection (cooperative jamming), we feel that there are some misleading issues related to their proposed secrecy encoding method. This why we re-formulate the *collective-secrecy* [15] encoding method in the current paper.

The encoding method of [15] uses a separate secret message encoding for each user, much like our own encoding scheme. However, unlike the present paper, the secrecy encoding of [15] employs a "superposition encoding scheme" (see Section III of both [15] and [16]). In the following paragraphs, we provide a brief description of this technique.

Take one user with power constraint $P$. The user generates two independent codebooks, in the following manner: the first codebook contains $2^{NR_s}$ $N$-dimensional codewords, and each letter of each codeword is independently generated, according to the realization of a Gaussian random variable of zero mean and variance $\alpha P$; the second codebook contains $2^{NR_0}$ $N$-dimensional codewords, and each letter of each codeword is independently generated, according to the realization of a Gaussian random variable of zero mean and variance $(1-\alpha)P$. The secret message – transmitted at rate $R_s$ – picks a codeword from the first codebook, while another codeword is randomly picked from the second codebook. The message transmitted by this user is the summation of the two codewords.

At a first glance, it appears that the transmitted message belongs to a codebook of $2^{N(R_s+R_0)}$ $N$-dimensional codewords, in which each letter of each codeword is the realization of a Gaussian random variable of variance $P$. Moreover, the codebook is already binned, like in Wyner's scheme [1], [17].

However, if the transmitted message is completely decodable by Bob, the rates $R_s$ and $R_0$ should be situated within the corresponding MAC rate region. For example, if we had a Gaussian eavesdropper channel where the AWGN variances were 1 for both channels, while the absolute squared channel coefficients are 1 for the main channel and $h_k$ for the eavesdropper's channel, the rates should satisfy $R_s \leq \log[1 + \alpha P]$, $R_0 \leq \log[1 + (1-\alpha)P]$, and $R_s + R_0 \leq \log[1 + P]$. But the first two conditions do not appear in [15].

Even if these conditions were satisfied, we believe that the "superposition encoding scheme" of [15] is not equivalent to Wyner's scheme. The key to Wyner's scheme is that each bin makes a "good" codebook for the eavesdropper. That is, given the secret key and the eavesdropper's received message, the bin chosen by the secret key conveys information to the eavesdropper at a rate arbitrarily close to the eavesdropper's channel capacity.

For the same toy model as above, the rate of each bin should be arbitrarily close to $\log[1 + Ph_k]$. However, under the "superposition encoding scheme" of [15], this rate cannot exceed $\log[1 + \alpha Ph_k]^3$. To achieve the capacity of the eavesdropper's channel, $\alpha$ would need to be arbitrarily close to 1. But then the codebook associated with the secret message would be generated with arbitrarily small power. If a positive secrecy rate $R_s$ is still desired, the intelligibility of the secret message at the legitimate receiver is compromised. Thus, we do not expect that the proposed encoding method of [15] will achieve the secrecy rates claimed therein.

---

[3]Note that although the second codebook has a rate equal to $\log[1 + Ph_k]$ in [15], this rate is not sustainable by the eavesdropper's channel with power constraint $\alpha P$.

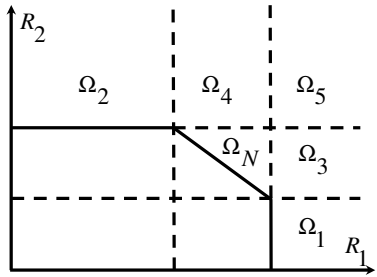

Fig. 11. The capacity region of a MAC.

At this point, we want to emphasize the fact that, except for the encoding method, the results of [15] and [16] are correct. However, our Theorem 4 is quite different than these results. Our secrecy-encoding scheme involves in the generation of the secret key the levels that are non-decodable by either Eve or Bob. In the context of [15] and [16], this would be equivalent to having some users transmit at rates not supported by the receiver, but still help with the transmission of secrecy. Although this approach would not make much sense in [15] and [16], it fits perfectly with the constraints of our compound-channel-like scenario.

## APPENDIX C
## PROOF AND MOTIVATION OF PROPOSITION 6

As we stated earlier, from Eve's point of view, the different encoding levels are very similar to different users. Therefore, Eve's channel can be seen as a multiple access channel (MAC), with $n$ users, each with a different power, but all sharing the same channel coefficient. However, to the best of our knowledge, in the current literature there is no treatment of the achievable rate region for a set of users when the other users are not decodable.

To motivate Proposition 6, we look at the two-user Gaussian MAC, the capacity region of which is given in Figure 11. Let the capacity of the first user's channel (when user 2 is absent) be $C_1 = \log(1+P_1/\sigma_N^2)$, and the capacity of the second user's channel (when user 1 is absent) be $C_2 = \log(1+P_2/\sigma_N^2)$. We know that the achievable rate region is given by all pairs $(R_1, R_2)$ that satisfy [18] $R_1 \leq C_1$, $R_2 \leq C_2$ and $R_1 + R_2 \leq \log(1+(P_1+P_2)/\sigma_N^2)$. This implies that when user 2 transmits at a rate $R_2 = C_2$, user 1 should be decoded by treating the second user as white Gaussian noise, and by performing successive interference cancellation. The first user's maximum decodable transmission rate is then $R_1 = \log(1+P_1/(\sigma_N^2+P_2))$.

However, it is not straightforward to see whether, when the second user uses a randomly generated Gaussian codebook at a rate $R_2 > C_2$ and cannot be decoded, the first user may employ a transmission rate larger than $\log(1+P_1/(\sigma_N^2+P_2))$ (region $\Omega_4$) . To justify our question, consider the following "decoding" method. First, a *list* of possible codewords is computed for user 2, by treating user 1 as interference, and selecting only those codewords of the second user's codebook that have a non-zero a posteriori probability. This list may be shorter than the second user's whole codebook, and the a posteriori probability of the codewords therein may be non-uniform. Then, using this information about user 2, we attempt decoding for user 1. Proposition 6 states that this method is no better than the one which treats user 2 as white noise.

In the remainder of this appendix we formulate and prove the following lemma, which considers a general Gaussian MAC, and from which the proof of Proposition 6 is straightforward to derive.

*Lemma 8:* Consider a Gaussian MAC with users $j \in \mathscr{I}$, each with average power constraint $P_j$, $j \in \mathscr{I}$, and each transmitting at a rate $R_j$, $j \in \mathscr{I}$. Let the variance of the additive white complex Gaussian noise be $\sigma_N^2$. We are interested in the maximum number of users that are decodable by the receiver. The maximal set of indices $\mathscr{I}_e \subseteq \mathscr{I}$ corresponding to these users that are perfectly decodable by the receiver is the largest of the sets $\mathscr{V}_e$ for which

$$\sum_{j \in \mathscr{S}} R_j \leq \log\big(1 + \frac{\sum_{j \in \mathscr{S}} P_j}{\sigma_N^2 + \sum_{i \in \mathscr{V}_{ne}} P_i}\big), \ \forall \mathscr{S} \subseteq \mathscr{V}_e, \quad (57)$$

where $\mathscr{V}_{ne}$ is the complement of $\mathscr{V}_e$ with respect to $\mathscr{I}$

*Proof:* Denote the largest of the sets $\mathscr{V}_e$ for which (57) holds by $\mathscr{V}_e^*$, and its complement in $\mathscr{I}$ by $\mathscr{V}_{ne}^*$. Assume that $\mathscr{V}_{ne}^*$ is non-empty (if it is empty, the result of the lemma is trivial). Note that all users in $\mathscr{V}_e^*$ are decodable. However, if all users of $\mathscr{V}_{ne}^*$ were decodable, then the property in (57) would hold for $\mathscr{I}$, hence $\mathscr{V}_e^*$ would no longer be the largest set with that property. Therefore, not all users of $\mathscr{V}_{ne}^*$ are decodable.

For the users in $\mathscr{V}_{ne}^*$, consider the the rate-region where: (1) none of the users can be decoded by treating the others as interference, (2) if any one user were decodable (i.e. provided to the receiver by a genie), then all other users would be decodable and (3) not all users are decodable. This region can be characterized as

$$\Omega_N = \Big\{ (R_i)_{i \in \mathscr{V}_{ne}^*} | R_j > \log\big(1 + \frac{P_j}{\sigma_N^2 + \sum_{k \in \mathscr{V}_{ne}^* \setminus \{j\}} P_k}\big),$$

$$\forall j \in \mathscr{V}_{ne}^*, \sum_{j \in \mathscr{S}} R_j \leq \log\big(1 + \frac{\sum_{j \in \mathscr{S}} P_j}{\sigma_N^2}\big), \ \forall \mathscr{S} \subset \mathscr{V}_{ne}^* \setminus \{i\},$$

$$\forall i \in \mathscr{V}_{ne}^*, \sum_{j \in \mathscr{V}_{ne}^*} R_j > \log\big(1 + \frac{\sum_{j \in \mathscr{V}_{ne}^*} P_j}{\sigma_N^2}\big) \Big\}$$

(for $|\mathscr{V}_{ne}^*| = 2$, $\Omega_N$ is represented in Figure 11).

Next we prove that such a rate region is non-empty for the users in $\mathscr{V}_{ne}^*$. To accomplish this, we show by mathematical induction that

$$\sum_{j \in \mathscr{S}} \log\big(1 + \frac{P_j}{\sigma_N^2 + \sum_{k \in \mathscr{V}_{ne}^* \setminus \{j\}} P_k}\big) \leq$$

$$\leq \log\big(1 + \frac{\sum_{j \in \mathscr{S}} P_j}{\sigma_N^2}\big), \ \forall \mathscr{S} \subset \mathscr{V}_{ne}^*, \quad (58)$$

which means that if the first condition in the definition of $\Omega_N$ holds for all $j \in \mathscr{S}$, this does not prevent the second condition in the definition of $\Omega_N$ from holding as well.

For $\mathscr{S} = \{i\}$ we have

$$\log\big(1 + \frac{P_i}{\sigma_N^2 + \sum_{j \in \mathscr{V}_{ne}^* \setminus \{i\}} P_j}\big) < \log\big(1 + \frac{P_i}{\sigma_N^2}\big). \quad (59)$$

If (58) holds for a set $\mathscr{S}$, that is

$$\sum_{j \in \mathscr{S}} \log \big(1 + \frac{P_j}{\sigma_N^2 + \sum_{k \in \mathscr{V}_{ne}^* \setminus \{j\}} P_k}\big) \leq$$
$$\leq \log \big(1 + \frac{\sum_{j \in \mathscr{S}} P_j}{\sigma_N^2}\big), \quad (60)$$

then adding another user $r$ to $\mathscr{S}$ we get

$$\sum_{j \in \mathscr{S} \bigcup \{r\}} \log \big(1 + \frac{P_j}{\sigma_N^2 + \sum_{k \in \mathscr{V}_{ne}^* \setminus \{j\}} P_k}\big) \leq$$
$$\leq \log \big(1 + \frac{\sum_{j \in \mathscr{S}} P_j}{\sigma_N^2}\big) +$$
$$+ \log \big(1 + \frac{P_r}{\sigma_N^2 + \sum_{j \in \mathscr{V}_{ne}^* \setminus \{r\}} P_j}\big) =$$
$$= \log \big(\frac{\sum_{j \in \mathscr{S}} P_j + \sigma_N^2}{\sigma_N^2} \cdot \frac{\sigma_N^2 + \sum_{k \in \mathscr{V}_{ne}^*} P_k}{\sigma_N^2 + \sum_{k \in \mathscr{V}_{ne}^* \setminus \{r\}} P_k}\big) \leq$$
$$\leq \log \big(1 + \frac{\sum_{j \in \mathscr{S} \bigcup \{r\}} P_j}{\sigma_N^2}\big), \quad (61)$$

where the last inequality holds because it is equivalent to the inequality

$$\frac{\sigma_N^2 + \sum_{k \in \mathscr{V}_{ne}^*} P_k}{\sigma_N^2 + \sum_{k \in \mathscr{V}_{ne}^* \setminus \{r\}} P_k} \leq \frac{\sigma_N^2 + \sum_{j \in \mathscr{S} \bigcup \{r\}} P_j}{\sigma_N^2 + \sum_{j \in \mathscr{S}} P_j}, \quad (62)$$

which is equivalent to the inequality

$$\Big[\sum_{k \in \mathscr{V}_{ne}^*} P_k\Big] \Big[\sum_{j \in \mathscr{S}} P_j\Big] \leq$$
$$\leq \Big[\sum_{k \in \mathscr{V}_{ne}^* \setminus \{r\}} P_k\Big] \Big[\sum_{j \in \mathscr{S} \bigcup \{r\}} P_j\Big], \quad (63)$$

which holds because $\sum_{k \in \mathscr{V}_{ne}^*} P_k - \sum_{j \in \mathscr{S}} P_j > P_r$.

At this point we know that $\Omega_N$ is a feasible (non-empty) rate region for the users in $\mathscr{V}_{ne}^*$ and (by construction) no user of $\Omega_N$ can be decodable. Note that the first condition in the definition of $\Omega_N$, i.e. that none of the users can be decoded by treating the others as interference, follows naturally for all users in $\mathscr{V}_{ne}^*$ (see the definition of $\mathscr{V}_e^*$ above), while the third condition, i.e. that not all users are decodable, is also characteristic of $\mathscr{V}_{ne}^*$, as we have already shown. Thus, the only restrictive condition on $\Omega_N$ is an upper bound on the rates (the second condition). In other words, any rate tuple which is feasible for the users in $\mathscr{V}_{ne}^*$ can be obtained from a rate tuple in $\Omega_N$ by *increasing* some of the rates.

But since none of the users is decodable for a rate tuple in $\Omega_N$, it is not possible that any non-empty set of users suddenly becomes decodable as some of the transmission rates increase. Equivalently, for Figure 11, since none of the two users is decodable when they transmit at a rate pair in $\Omega_N$, it is not possible that one of them becomes decodable by increasing its rate (such that the rate pair moves to either $\Omega_3$ or $\Omega_4$ or $\Omega_5$).

To see this, using the notation already established in the previous sections ($\mathbf{X}_\mathscr{S}$ is the set of transmitted sequences of all users in $\mathscr{S}$, $\mathbf{Z}$ is the received sequence, $\mathbf{Q}$ is the noise sequence), for any set $\mathscr{S} \subset \mathscr{V}_{ne}^*$ we can write

$$H(\mathbf{X}_\mathscr{S} | \mathbf{Z}, \mathbf{X}_{\mathscr{V}_e^*}) =$$
$$= H(\mathbf{X}_\mathscr{S} | \mathbf{X}_{\mathscr{V}_e^*}) + H(\mathbf{Z} | \mathbf{X}_\mathscr{S}, \mathbf{X}_{\mathscr{V}_e^*}) - H(\mathbf{Z} | \mathbf{X}_{\mathscr{V}_e^*}) =$$
$$= H(\mathbf{X}_\mathscr{S}) + H(\mathbf{Z} | \mathbf{X}_\mathscr{S}, \mathbf{X}_{\mathscr{V}_e^*}) - H(\mathbf{Z} | \mathbf{X}_{\mathscr{V}_e^*}) =$$
$$= \sum_{j \in \mathscr{S}} R_j + H\big(\sum_{i \in \mathscr{V}_{ne}^* \setminus \mathscr{S}} \mathbf{X}_j + \mathbf{Q}\big) - H(\mathbf{Z} | \mathbf{X}_{\mathscr{V}_e^*}). \quad (64)$$

where the last two equalities follow from the independence of the users. Note that $H(\sum_{i \in \mathscr{V}_{ne}^* \setminus \mathscr{S}} \mathbf{X}_j + \mathbf{Q})$ increases with any rate $R_i$, $i \in \mathscr{V}_{ne}^* \setminus \mathscr{S}$. But since all rate tuples under consideration are outside the capacity region of the users in $\mathscr{V}_{ne}^*$, the received sequence $\mathbf{Z}$ displays an asymptotic equipartition property, as noticed in [19]. Intuitively, this happens because, as the rate tuple moves outside the capacity region, the volume of the typical set of received sequences becomes as large as the volume of the whole channel output space. The immediate implication is that outside the capacity region $H(\mathbf{Z} | \mathbf{X}_{\mathscr{V}_e^*})$ is a constant function of the rates of the users in $\mathscr{V}_{ne}^*$.

This concludes our proof that the entropy of the users in any subset of $\mathscr{V}_{ne}^*$ increases as any of the rates increase. But since any rate tuple for the users in $\mathscr{V}_{ne}^*$, situated outside of the rate region $\Omega_N$, can be obtained from a tuple inside $\Omega_N$, by increasing at least one of the rates, and since none of the users in $\Omega_N$ is decodable, this implies that none of the users in $\mathscr{V}_{ne}^*$ is decodable. ∎

A notable consequence of Lemma 8 above is that, in a MAC scenario employing Gaussian coding, whenever a user transmits at a rate which exceeds its channel capacity, the best strategy for the other users is to treat it as noise.

## REFERENCES

[1] A. D. Wyner, "The wire-tap channel," *The Bell System Technical Journal*, vol. 54, pp. 1355–1387, Oct. 1975.
[2] I. Csiszar and J. Korner, "Broadcast channels with confidential messages," *IEEE Trans. Inform. Theory*, vol. 24, pp. 339–348, May 1978.
[3] Z. Li, R. Yates, and W. Trappe, "Secret communication with a fading eavesdropper channel," *Proc. IEEE Int. Symp. on Inform. Theory (ISIT)*, June 2007.
[4] P. K. Gopala, L. Lai, and H. E. Gamal, "On the secrecy capacity of fading channels," *IEEE Trans. Inform. Theory*, vol. 54, pp. 4687–4698, Oct. 2008.
[5] Y. Liang, H. V. Poor, and S. Shamai, "Secure communication over fading channels," *IEEE Trans. Inform. Theory*, vol. 54, pp. 2470–2492, June 2008.
[6] A. Khisti, A. Tchamkerten, and G. Wornell, "Secure broadcasting over fading channels," *IEEE Trans. Inform. Theory*, vol. 54, pp. 2453–2469, June 2008.
[7] A. J. Goldsmith and P. P. Varaiya, "Capacity of fading channels with channel state information," *IEEE Trans. Inform. Theory*, vol. 43, pp. 1986–1992, Nov. 1997.
[8] S. Shafiee and S. Ulukus, "Capacity of multiple access channels with correlated jamming," *Military Communications Conference, MILCOM*, vol. 1, pp. 218–224, Oct. 2005.
[9] A. Kashyap, T. Basar, and R. Srikant, "Correlated jamming on mimo Gaussian fading channels," *IEEE Trans. Inform. Theory*, vol. 50, pp. 2119–2123, Sept. 2004.
[10] E. Altman, K. Avrachenkov, and A. Garnaev, "A jamming game in wireless networks with transmission cost," *Proceedings of Net-Coop, Avignon, France*, June 2007.
[11] S. N. Diggavi and T. Cover, "The worst additive noise under a covariance constraint," *IEEE Trans. Inform. Theory*, vol. 47, pp. 3072–3081, Nov. 2001.
[12] G. T. Amariucai and S. Wei, "Active eavesdropping in fast fading channels," *Proc. IEEE Military Commun. Conf. (MILCOM)*, Oct. 2009.

[13] T. M. Cover, "Broadcast channels," *IEEE Trans. Inform. Theory*, vol. 18, pp. 2–14, Jan. 1972.
[14] B. Schneier, *Applied cryptography*. John Wiley & Sons, 1996.
[15] E. Tekin and A. Yener, "The general Gaussian multiple-access and two-way channels: achievable rates and cooperative jamming," *IEEE Trans. Inform. Theory*, vol. 54, pp. 2735–2751, June 2008.
[16] ——, "The Gaussian multiple access wire-tap channel," *IEEE Trans. Inform. Theory*, vol. 54, pp. 5747–5755, Dec. 2008.
[17] S. K. Leung-Yan-Cheong and M. E. Hellman, "The Gaussian wire-tap channel," *IEEE Trans. Inform. Theory*, vol. 24, pp. 451–456, July 1978.
[18] T. M. Cover and J. A. Thomas, *Elements of information theory (second ed.)*. Hoboken, New Jersey: John Wiley & Sons, Inc., 2006.
[19] X. Wu and L.-L. Xie, "Asymptotic equipartition property of output when rate is above capacity," *arXiv:0908.4445v1 [cs.IT]*, Aug. 2009.